# Electronic Energy Migration in Microtubules

Aarat P. Kalra [a], Alfy Benny [a], Sophie M. Travis [b], Eric A. Zizzi [c], Austin Morales-Sanchez [a], Daniel G. Oblinsky [a], Travis J. A. Craddock [d], Stuart R. Hameroff [e], M. Bruce MacIver [f], Jack A. Tuszyński [c, g, h], Sabine Petry [b], Roger Penrose [i], Gregory D. Scholes [a, *]

[a] Department of Chemistry, New Frick Chemistry Building, Princeton University, NJ 08544, USA
[b] Department of Molecular Biology, Schultz Laboratory, Princeton University, NJ 08544, USA
[c] Department of Mechanical and Aerospace Engineering (DIMEAS), Torino 10129, Italy
[d] Departments of Psychology & Neuroscience, Computer Science, and Clinical Immunology, Nova Southeastern University, Ft. Lauderdale, FL 33314, USA
[e] Department of Anesthesiology, Center for Consciousness Studies, University of Arizona, Tucson, Arizona, USA
[f] Department of Anesthesiology, Stanford University School of Medicine, Stanford, CA 94305, USA
[g] Department of Physics, University of Alberta, Edmonton, Alberta T6G 2E1, Canada
[h] Department of Oncology, University of Alberta, Edmonton, Alberta T6G 1Z2, Canada
[i] Mathematical Institute, Andrew Wiles Building, University of Oxford, Radcliffe Observatory Quarter, Woodstock Road, Oxford, OX2 6GG, United Kingdom

[*] To whom correspondence should be addressed: gscholes@princeton.edu

## Abstract

The repeating arrangement of tubulin dimers confers great mechanical strength to microtubules, which are used as scaffolds for intracellular macromolecular transport in cells and exploited in biohybrid devices. The crystalline order in a microtubule, with lattice constants short enough to allow energy transfer between amino acid chromophores, is similar to synthetic structures designed for light harvesting. After photoexcitation, can these amino acid chromophores transfer excitation energy along the microtubule like a natural or artificial light-harvesting system? Here, we use tryptophan autofluorescence lifetimes to probe inter-tryptophan energy hopping in tubulin and microtubules. By studying how quencher concentration alters tryptophan autofluorescence lifetimes, we demonstrate that electronic energy can diffuse over 6.6 nm in microtubules. We discover that while diffusion lengths are influenced by tubulin polymerization state (free tubulin *versus* tubulin in the microtubule lattice), they are not significantly altered by the average number of protofilaments (13 *versus* 14). We also demonstrate that the presence of the anesthetics etomidate and isoflurane reduce exciton diffusion. Energy transport as explained by conventional Förster theory (accommodating for interactions between tryptophan and tyrosine residues) does not sufficiently explain our observations. Our studies indicate that microtubules are, unexpectedly, effective light harvesters.

Microtubules | electronic energy transfer | light harvesting

> **Significance Statement**
> It is well known that aromatic amino acids in proteins absorb ultraviolet light, but it is generally assumed that proteins are not effective light-harvesters. Here, we study light harvesting by microtubules, which are micrometer-length tubular protein assemblies, ubiquitous in cells. Our experiments reveal that energy can migrate by diffusive energy transfer over unexpectedly large distances (6.6 nm). We find that conventional Förster theory predicts a diffusion length of only ~2.3 nm; insufficient to explain our observations. Introducing the anesthetics etomidate and isoflurane decreases the observed energy diffusion length. We conclude that it is worth considering protein assemblies, like microtubules, for ultraviolet light-harvesting systems.

## Introduction

Microtubules are cylindrical polymers of the protein α, β tubulin that play a variety of structural roles in the cell. Microtubules facilitate chromosome segregation during mitosis, generate intracellular forces, form a 'railroad network' for macromolecular transport and provide mechanical support for organelle positioning (1, 2). Tubulin is packed in an exquisitely systematic manner in a microtubule, ordered both vertically in columns called protofilaments, and horizontally in helical rings (Fig. 1A-C; (3)). The emergence of a lattice (which, despite the nanometer-scale dimensions of constituent tubulin dimers, can stretch unbroken across several hundreds of micrometers (4)) and the hollow cross section of a microtubule confer structural rigidity, allowing robust execution of a variety of mechanical tasks (5). Beyond the cell, these properties have allowed the widespread use of microtubules scaffolds in transport-based nanodevices (6-10), wherein motor proteins transport macromolecular cargo across large distances up to the millimeter range by 'walking' along the microtubule lattice (11).

It is well known that inter-aromatic residue energy transfer over small distances takes place in proteins (12-14). Compared to pigments that absorb light in the visible region of the spectrum, aromatic residues absorb and emit UV light, have low molar extinction coefficients and small photoluminescence quantum yields. Consequently, proteins are not optimal light-harvesting systems for long-range photoexcitation energy migration (15). The lattice-type structure of microtubules that incorporates periodic arrays of aromatic amino acids, however, is qualitatively similar to the structure of molecular aggregates (16-19) and light-harvesting complexes like phycobilisomes (20, 21), potentially enabling excitation energy transport over long distances (22). How long are these distances?

Here, we exploit tryptophan autofluorescence to show that the 2D photoexcitation diffusion length in microtubules (6.6 ± 0.1 nm) is substantially higher than that predicted by conventional Förster theory (1.66 nm for inter-tryptophan interactions and 2.27 nm including tyrosine-tryptophan interactions). The diffusion length for electronic energy migration in microtubules is thus close to the size of tubulin dimer, comparable to that reported in some photosynthetic complexes (23-26). Changing the number of protofilaments from 13 to 14 has a negligible effect on the 2D photoexcitation diffusion length. While a high background signal made it difficult to quantify signal from vinblastine-induced tubulin oligomers, our data also suggest that photoexcitation diffusion lengths decrease in tubulin oligomers (formed in the presence of vinblastine), indicating that photoexcitation may migrate more effectively along protofilaments, as opposed to rings. Notably, we find that the presence of the anesthetics etomidate and isoflurane lowered photoexcitation diffusion coefficients (and thus diffusion lengths) in microtubules. Considering the factors (enumerated above) that make tryptophan a far-from-optimal chromophore, our results demonstrate that electronic energy migration is surprisingly efficient in microtubules.

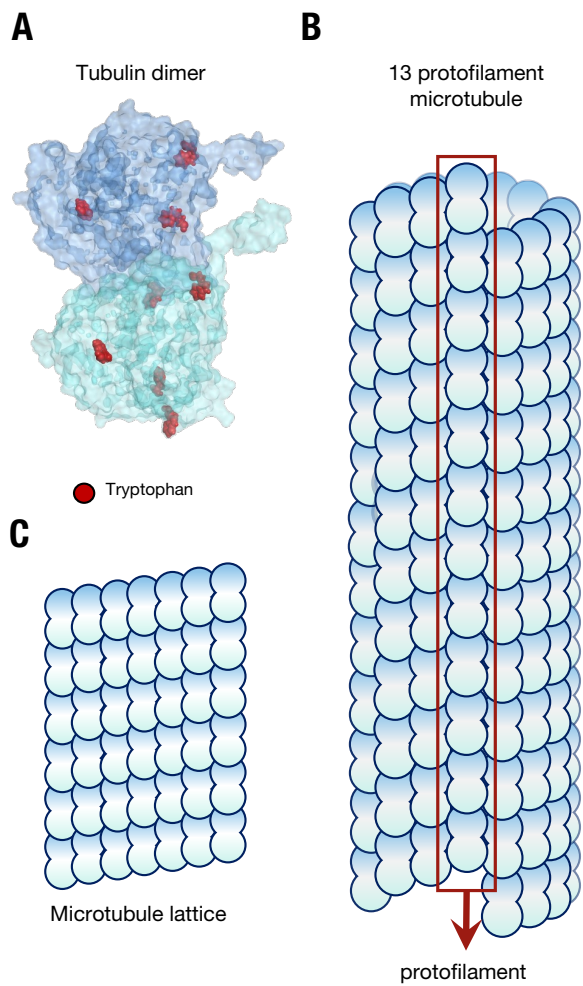

**Figure 1. The structure of microtubules from a lattice of tubulin.** (**A**), The tubulin dimer with tryptophan residues marked in red, the C-termini 'tails' can be seen protruding from each monomer. (**B**), the structure of a microtubule, showing constituent arrangement of tubulin dimers, and the presence of a 'seam' (**C**) the repeating 'lattice' of tubulin dimers in a microtubule.

## Results

**Tryptophan quenching rates can distinguish between free GTP-tubulin and microtubules.** We introduced an externally conjugated tryptophan fluorescence quencher, AMCA (7-amino-4-methyl coumarin-3-acetic acid; Fig. 2A) to study inter-tryptophan hopping in microtubules (SI Appendix, Fig. 3E). We found that tryptophan fluorescence lifetimes in microtubules shortened on increasing AMCA concentration (Fig. 4A) and, consistent with fluorescence quenching through electronic energy transfer, also observed AMCA fluorescence emission (Fig. 2E, F) on tryptophan photoexcitation in steady-state experiments. To quantify energy transfer between tryptophan and AMCA, we used the Stern-Volmer equation for static quenching, $(\frac{\tau_0}{\tau} = 1 + k_Q[Q]\tau_0)$ where $\tau$ is the tryptophan lifetime at AMCA concentration $[Q]$, $\tau_0$ is the tryptophan lifetime in the absence of AMCA, and $k_Q$ is the rate constant for quenching by electronic energy transfer (Fig. 5A). Our analysis for microtubules and free GTP-tubulin in solution revealed quenching constants of 23.8 ± 0.8 ns$^{-1}$ and 7.5 ± 0.4 ps$^{-1}$, respectively (Fig. 4A, Table 1). AMCA itself contributed negligible signal over the relevant wavelength range (Fig. S6). The higher quenching rate in microtubules can be explained by sensitization of AMCA quenching sites remote from the photo-excited tubulin dimer, which can occur by multiple inter-tryptophan energy transfer steps (energy migration).

To analyze our experimental results, we estimated photoexcitation diffusion lengths using Stern-Volmer analysis (27). The measured lifetimes $\tau$ and $\tau_0$ in the Stern-Volmer equation are related to the diffusion coefficient $D$ as shown in equation (1):

$$\frac{1}{\tau} = \frac{1}{\tau_0} + 4\pi r D[Q] \qquad (1)$$

Here, $r$ is the sum of the tryptophan and AMCA radii (assumed here to be 1 nm). The diffusion coefficient can be used to calculate the photoexcitation diffusion length $L$ over the time duration $\tau$ to which a photoexcitation migrates on a microtubule before being quenched by AMCA as shown in equation (2).

$$L = \sqrt{4D\tau} \qquad (2)$$

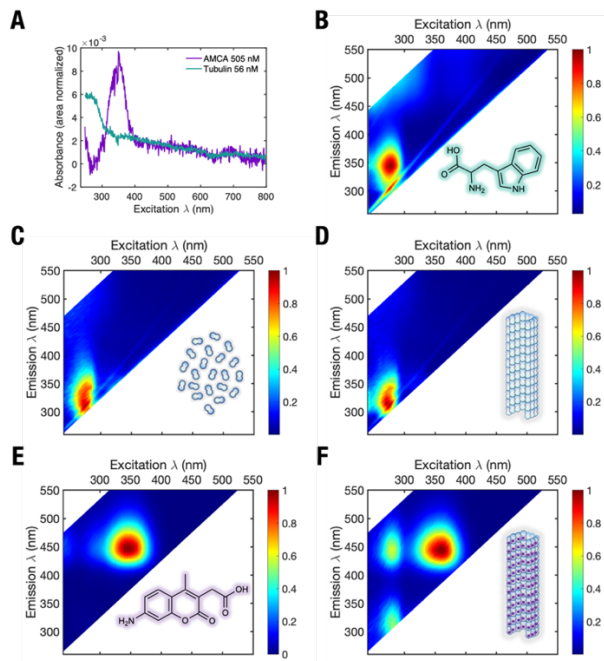

**Figure 2. Steady-state spectra of tubulin and microtubules (*A*),** Absorbance spectra of free tubulin (teal) and free AMCA (purple) in solution. Intensity normalized fluorescence spectra of **(*B*)**, free DL tryptophan in solution showing highest fluorescence emission at excitation wavelengths between 270 and 300 nm. **(*C*)** unpolymerized GTP-tubulin, **(*D*)**, microtubules polymerized using GTP-tubulin, **(*E*)**, free AMCA and **(*F*)**, microtubules polymerized using AMCA-labeled GTP-tubulin. An energy transfer peak not observed in Fig. 1A and 1B (at excitation 280-300 nm, and emission at 420-450 nm) is clearly visible. Colors represent photoluminescence intensity.

Our data yielded a 2D diffusion coefficient of 3.15 ± 0.1 × 10$^{-5}$ cm$^2$.s$^{-1}$ and a corresponding diffusion length of 6.64 ± 0.1 nm for microtubules (Table 1; Fig. 5B, C). Given that a single tubulin dimer occupies the same volume as a sphere of diameter 7.4 nm (from its dimensions of 4 nm × 6.5 nm ×8 nm), a diffusion length of 6.64 nm suggests that energy transfer between tryptophan residues across a single tubulin dimer can occur in a microtubule. Depending on the locations of the residue where the initial photoexcitation occurred, it could reasonably migrate to the adjacent tubulin dimer within the microtubule lattice.

We simulated (incoherent) energy migration in microtubules using the 31-tubulin long microtubule crystal structure (SI Appendix, Fig. 6A). The primary pathway for energy migration is tryptophan to tryptophan. However, given its spectral overlap with tryptophan (12), tyrosine could serve as an activated intermediate. While the spectral overlap for tryptophan-tyrosine (uphill) energy transfer is small and tryptophan has a higher transition dipole transition moment ($\mu_{TYR} = 1.356\ D$, $\mu_{TRP} = 2.77\ D$ (SI Appendix), these factors might be offset by the large number of tyrosine residues compared to tryptophan (34 tyrosine residues *versus* 8 tryptophan residues in the tubulin sequence 1JFF (28)) . We accounted for both tyrosine and tryptophan chromophores in the simulations.

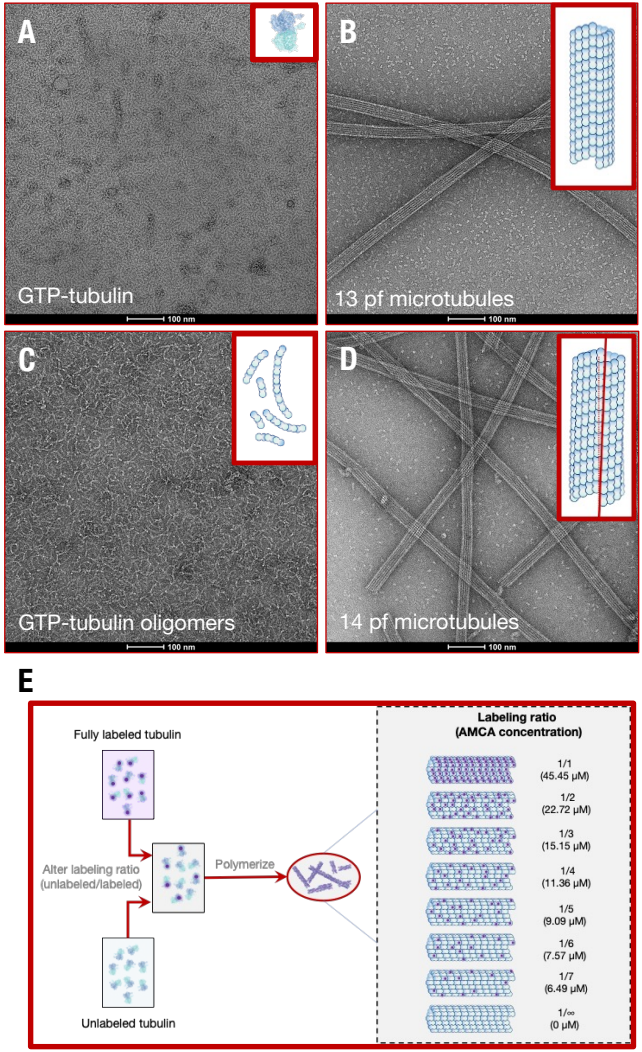

**Figure 3. Confirmation and polymerization of tubulin polymorphs.** Negative stain electron microscopy of tubulin polymorphic geometries for **(A)**, Free GTP-tubulin in solution, **(B)**, GTP-tubulin polymerized 13 protofilament microtubules, **(C)**, Free GTP-tubulin oligomers polymerized using 100 μM vinblastine in solution and **(D)**, GMPCPP-tubulin polymerized 14 protofilament microtubules. Insets show schematic of tubulin polymer structure. Methodology used to perform negative staining is described in SI Appendix. **(E)**, Routine to polymerize microtubules with different AMCA labeled tubulin: unlabeled tubulin ratios, thus different AMCA concentrations. See SI Appendix for protocols used to assemble different tubulin polymerization states and perform electron microscopy.

We calculated the coupling strengths for *homo* (between same residues) and *hetero* (between tyrosine and tryptophan) energy transfer using the ideal dipole approximation (29):

$$V_{Coul}^{ij} = \frac{1}{4\pi\epsilon_0} \frac{k|\mu_i||\mu_j|}{n^2\ r^3} \qquad (3)$$

Here, $\mu_i$ and $\mu_j$ denote the transition dipole moments of the $i^{th}$ and $j^{th}$ residues, $\kappa$ is the orientation factor, $r$ is the separation distance, $\epsilon_0$ is the permittivity of free space and $n$ is the refractive index of the medium surrounding the microtubule (assumed to be $n = 1.4$). The hetero-molecular coupling between tyrosine and tryptophan residues shows the highest contribution to the $V_{Coul}$, followed by tyrosine-tyrosine coupling strengths. tryptophan-tryptophan coupling strengths are the lowest, as a result of longer intermolecular distances between them (Fig. 7B, C).

We calculated the diffusion lengths on the microtubule crystal structure using a kinetic Monte Carlo algorithm (SI Appendix, Fig. S14). The FRET rate ($k_{FRET}$) between residues $i$ and $j$ was calculated using equation (4) (30):

$$k_{FRET}^{ij} = \frac{2\pi}{\hbar}|V_{Coul}^{ij}|^2 J_{SO} \qquad (4)$$

Absorbance and emission spectra accommodating the tryptophan peak wavelength to our experimental observations were used to calculate the spectral overlap ($J_{SO}$) (31) by:

$$J_{SO} = \int_0^\infty f(\epsilon)a(\epsilon)d\epsilon \qquad (5)$$

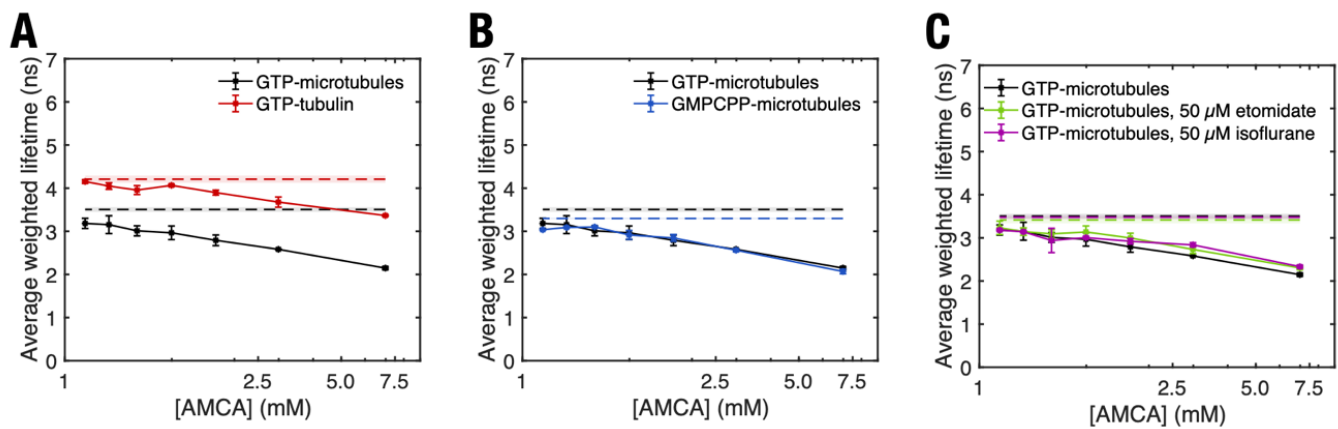

**Figure 4. Average weighted tryptophan fluorescence lifetime of different tubulin polymerization state as a function of AMCA concentration.** (**A**), Free GTP tubulin compared to that of microtubules polymerized using GTP-tubulin, (**B**), microtubules polymerized using GTP-tubulin to that of those polymerized using GMPCPP-tubulin, (**C**), microtubules polymerized using GTP-tubulin in the presence and absence of etomidate and isoflurane. Dashed lines and shaded region represent mean and standard deviation of tryptophan lifetimes in different tubulin polymerization states in the absence of AMCA. p-values were calculated to determine significance of differences in average weighted lifetimes between GTP microtubules and other tubulin polymerization states (see Fig. S10). Error bars represent standard deviation of experiments conducted $n = 3$ to $n = 5$ times.

Here, $f(\epsilon)$ and $a(\epsilon)$ are the emission and absorption spectra of the donor and acceptor residues,

respectively, which are area normalized on an energy scale (SI Appendix, Table S11).

We simulated 3000 trajectories, first considering only inter-tyrosine interactions and then considering tyrosine-tryptophan interactions. In both simulations, the obtained diffusion lengths (1.66 nm and 2.27 nm, respectively) are appreciably lower that the diffusion lengths obtained experimentally. A possible explanation could be that some of the interchromophore separations are too close for the dipole approximation to be accurate. It is even possible that short-range contributions to the electronic coupling arising from intermolecular orbital overlap may be significant, as has been found te be the case in other tightly packed aggregates (32-34). To obtain a rough estimate of the extent of short-range effects, we calculated the hole ($t_h$) and electron ($t_h$) transfer Integrals for five representative TYR dimers present in close proximity (inter-aromatic distance ($R_{IA}$) < 6.3 Å) (Table S12, Figure S15, Supporting Information)(35). We observed charge transfer Integral magnitudes ranging up to 419 cm$^{-1}$, which suggests that the intermolecular orbital overlap mediated interactions could increase the electronic coupling, and therefore the predicted exciton diffusion length.

**Electronic energy migration is dampened by anesthetics etomidate and isoflurane.** While microtubules have been hypothesized to support long-range dipole-switching of aromatic residues for information processing roles in neurons (36, 37), experimental support has not yet been provided. Anesthetics have been modelled to bind to tubulin in the microtubule interfering with such long-range interactions, and thereby inhibiting dipole-based information processing (38-40). An experimental evaluation of anesthetic action on long-range dipole switching in microtubules has also not yet been performed.

To investigate the influence of anesthetics on tryptophan fluorescence quenching, we introduced the anesthetics etomidate and isoflurane (41, 42) into our assay, and measured their effect on tryptophan fluorescence lifetimes. Introducing 50 µM etomidate and isoflurane significantly lowered tryptophan quenching by AMCA in GTP-polymerized microtubules (Fig. 4C). The presence of etomidate and isoflurane decreased the 2D diffusion lengths from 6.6 ± 0.1 nm to 5.6 ± 0.1 nm and 5.8 ± 0.2 nm, respectively (Fig. 5C). To investigate if this result could arise from the binding of etomidate and isoflurane to tubulin in the microtubule lattice, we built and refined computational models of human tubulin dimers and performed docking simulations (SI Appendix, Fig. S11).

Our results showed that both macromolecules exhibited affinities to 16 distinct binding sites (including the vinca, taxol and colchicine sites), but with low predicted binding energies, which indicate weak to moderate interactions with tubulin, with no distinct preference for any of these sites. Taken together, these findings suggest that etomidate and isoflurane 'dampen' energy transfer between tryptophan and AMCA due to changes in the dielectric screening of electronic couplings.

Notably, these anesthetics did not contribute to overall solution fluorescence nor the fluorescence spectra of free GTP-tubulin or AMCA (Fig. S6), demonstrating that they did not chemically interfere with either tryptophan or AMCA lifetime.

**Long-Range electronic energy migration also occurs in 14 protofilament microtubules.** Tubulin polymerization in the presence of GMPCPP, a slowly hydrolysable analog of GTP (43), results in the formation of a predominantly 14 protofilament microtubule with increased tryptophan-tryptophan distances within each tubulin dimer, as well as between neighboring dimers (44) (Fig. 3B). We asked if inter-tryptophan energy hopping would be significantly altered in 14 protofilament microtubules because of a difference in tryptophan nearest neighbor distances (Table S1). Our results showed that while unlabeled tryptophan fluorescence lifetimes in GMPCPP-microtubules were significantly lower than those in GTP-microtubules (p-values <0.005; Fig. S10B), the lifetime trend in the presence of AMCA in GMPCPP-microtubules resembled that of GTP- microtubules (p-values >0.005; Fig. 4B). Stern-Volmer analysis revealed different tryptophan quenching rates GMPCPP-microtubules arose from differences in tryptophan fluorescence lifetime in the absence of AMCA. Consequently, long-range electronic energy transfer was observed in GMPCPP-microtubules (diffusion lengths of 6.1 ± 0.1 nm; Fig. 5B), consistent with multiple step hopping. 14 protofilament microtubules are often used in microtubule-based mechanical devices because of their greater flexural rigidity than 13 protofilament microtubules (45, 46).

## Discussion

While the mechanical properties of microtubules and their interaction with molecular machines have been used as key components in transport-based nanodevices (6), their autofluorescence has not yet been exploited. In this work, we find that microtubules are more effective light-harvesters than anticipated for a protein devoid of bound chromophores.

We showed that tryptophan autofluorescence can distinguish between polymerized and unpolymerized tubulin. Stern-Volmer analysis revealed that the 2D diffusion length of tryptophan photoexcitation in microtubules is comparable to that observed in some photosynthetic complexes (23-25) and organic photovoltaic devices (47-49). Our observations suggest that photoexcitation diffusion takes place in a manner that is dependent on the tubulin polymorph type (tubulin dimers *versus* oligomers *versus* microtubules). Polymerizing into microtubules enhances diffusion lengths from 4 ± 0.1 nm in free GTP_tubulin to 6.64 ± 0.1 nm in microtubules. Subsequently adding tubulin-binding agents (such as anesthetics) decreases the observed diffusion length. The finding of long-range electronic energy transfer shows the versatility of tubulin-based polymers. Of course, it is unlikely that biological systems utilize the light-harvesting properties of tubulin, but it is fascinating that this protein macrostructure exhibits such photoexcitation diffusion lengths.

Photoexcitation diffusion in molecular systems is widely studied by a variety of techniques, often relying on exciton-exciton annihilation (17, 20, 21, 25, 27, 50-54). The two-dimensional tryptophan photoexcitation diffusion length of 6.64 nm is surprisingly high in microtubules, given that they are optimized to play structural roles in the cell, and that the equivalent value for chlorophyll *a*, which is optimized for electronic energy transfer, is only 20-80 nm (23-25, 55). Because tryptophan absorbs light poorly (molar extinction coefficient of ~5600 $M^{-1}$ $cm^{-1}$ (56)) compared to chlorophyll *a* (molar extinction coefficient of ~110,000 $M^{-1}$ $cm^{-1}$ (57)) photoexcitation diffusion on a lengths scale of several nanometers was not expected.

We also attempted to understand the preferred path of photoexcitation migration. It is likely that photoexcitation energy migration take place preferentially along a protofilament or along tubulin ring along a helical path (Fig. 7, Fig. S13). To determine if long-range photoexcitation migration could take place in the absence of lateral contacts between tubulin dimers, we polymerized GTP-tubulin in the presence of 100 µM vinblastine, which binds at axial interface between two tubulin dimers, forming tubulin oligomers (Fig. 3C; (58). Vinblastine produced a signal in our fluorescence lifetime measurement (Fig. S3B), making it difficult to quantify the tryptophan fluorescence because of overlapping fluorescence.

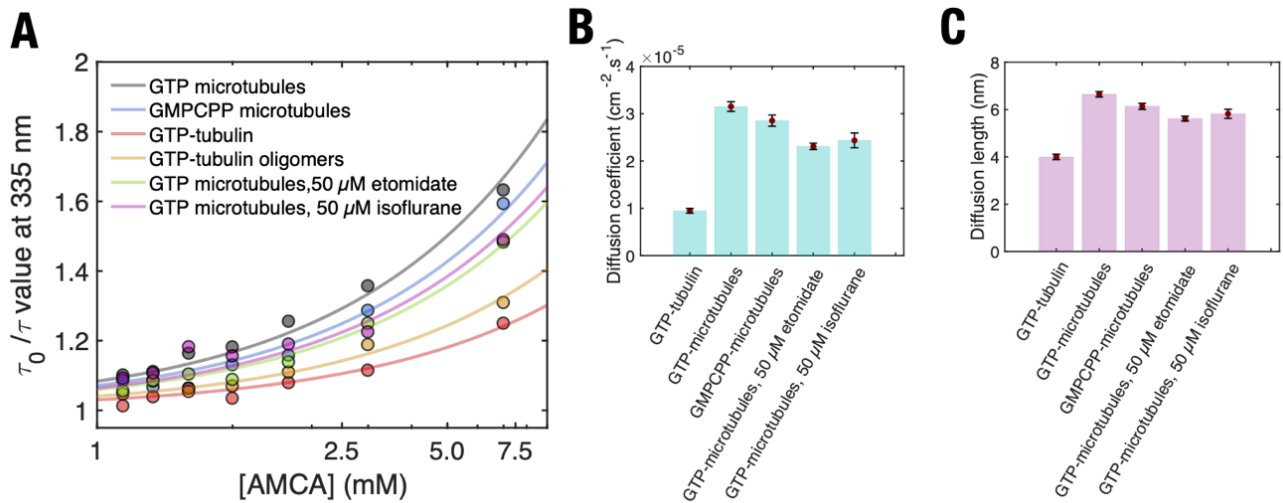

**Figure. 5. Parameters extracted from Stern-Volmer analysis to estimate diffusion parameters.** (**A**), Stern-Volmer plot for fitting different tubulin polymers with the static quenching model as shown in main text. Lines represent line of best fit to data points shown in scatter plot. (**B**), Diffusion coefficients and (**C**), diffusion lengths of tryptophan excitation in different tubulin polymerization states. Data shown in Table S5. The errors associated with (B) are standard errors for the fit to determine the diffusion coefficient. The error bar associated with (C) are calculated after propagating errors from experimental values included in equations 1 and 2.

Thus, after estimating the time resolved fluorescence decay of tubulin oligomers, we subtracted the short lifetime component (Fig. S8, S9), using only two lifetimes to determine the diffusion parameters of tryptophan (Fig. S7, Fig. 5A). Nevertheless, the diffusion length obtained from such analysis (4.6 ± 0.1 nm) indicates long-range electronic energy transport in oligomers. While these diffusion lengths are lower than those observed in microtubules, we speculate that longitudinal contacts between tubulin dimers, resembling those in a protofilament, are sufficient to induce long-range energy transport in tubulin polymers. The observation of diffusion lengths in GMPCPP microtubules also supports the notion that a repeating arrangement of tubulin dimers is sufficient to allow long-range photoexcitation energy transfer.

**Table 1.** $K_Q$ (mean ± standard error of fit) and Adjusted $R^2$ values for fitting different tubulin polymers with the static quenching model as shown in main text.

| Tubulin polymerization state | $K_Q$ (ns$^{-1}$) | Adj $R^2$ |
| --- | --- | --- |
| Free GTP-tubulin | 7.46 ± 0.4 | 0.96 |
| 13 protofilament microtubules (GTP-tubulin polymerized) | 23.84 ± 0.8 | 0.98 |
| 14 protofilament microtubules (GMPCPP-tubulin polymerized) | 21.59 ± 0.9 | 0.98 |
| 13 protofilament microtubules with 50 µM etomidate | 17.47 ± 0.5 | 0.99 |
| 13 protofilament microtubules with 50 µM isoflurane | 18.43 ± 1.2 | 0.91 |

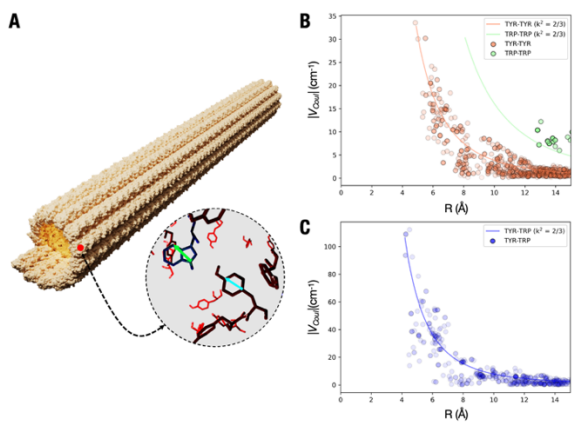

**Figure 6. Theoretical estimation of interactions among tryptophan and tyrosine residues.** *(A)*, Crystal structure of a microtubule composed of 31 tubulin dimers stacked vertically showing the dipole moment orientations of representative tryptophan (green; TRP) and tyrosine (cyan; TYR) residues. *(B)*, Distribution of the coupling constant $V_{Coul}$ between TYR-TYR, TRP-TRP and *(C)*, TYR-TRP residues in 31-dimer long microtubule crystal structure. A projection of $V_{Coul}$ with orientation factor ($\kappa^2$) of 2/3 is represented as solid lines.

It is also worth noting that introducing a convulsant (50 μM picrotoxin) also decreased tryptophan quenching by AMCA. These additional experiments, that we do not report here suggest that anesthetics are not the only macromolecules that decrease photoexcitation diffusion in microtubules. Thus, we anticipate that in addition to anesthetics, microtubule associated proteins (MAPs), microtubule associated drugs (MADs) and other tubulin interacting agents form a 'cytoexcitonic' framework for altering electronic energy migration in microtubule-based devices. Biochemically tuned electronic energy migration would allow microtubule architectures to be used as targets for photobiomodulation (9, 59), and to transduce photonic energy in biohybrid devices (60).

A key finding of our study is that photoexcitation diffusion cannot be explained *via* conventional Förster theory, which accounts for only dipole-dipole interactions between tryptophan and tyrosine residues. A satisfactory explanation of our observations may include solvent screening by the protein matrix (61), electronic coupling beyond the point dipole approximation in Förster theory (62, 63) and accommodation for molecular exciton states in tubulin (26). The tuning of photoexcitation diffusion length through biochemical parameters of microtubules (tubulin polymerization, introduction of anesthetics) was a further intriguing observation. Our work shows that protein polymers may be suitable for active materials in biologically sourced electronic devices where UV photoexcitation is desired.

### Materials and Methods

**Tubulin polymerization.** Tubulin was polymerized and stabilized as described in previous literature (10, 64). Briefly, porcine brain tubulin stock was prepared by reconstituting lyophilized tubulin powder (T240; Cytoskeleton Inc, Denver, CO, USA) in BRB80 buffer supplemented with 10% glycerol at 1 mM GTP. AMCA labeled tubulin solution was prepared by reconstituting lyophilized AMCA labeled tubulin powder (TL440m; Cytoskeleton Inc, Denver, CO, USA) in BRB80 supplemented with 10% glycerol and 1 mM GTP, to a final concentration of 45 μM tubulin. Labeling efficiency was 1-2 AMCA molecules per tubulin dimer, attached at random surface lysines. Different volumes of labeled tubulin solution were added to tubulin solution to prepare solutions containing different AMCA labeling ratios. These solutions were incubated for 45 minutes at 37 °C to polymerize microtubules and subsequently stabilized using 100 μM Paclitaxel (BRB80T) to ensure no microtubule depolymerization. Free GTP-tubulin stock in 100 μM vinblastine (BRB80V) for 45 minutes at 37 °C to stabilize GTP-tubulin oligomers. Measurements on free tubulin in solution were performed in the presence of 10 mM $CaCl_2$ (BRB80Ca) to ensure that free tubulin remained in solution.

**Steady-State Measurements.** The sample absorbance was measured as a function of light frequency using a UV-Vis spectrometer complemented with an integrating sphere (Cary 6000i, Agilent Technologies, Santa Clara, CA, USA). The absorption spectra were recorded using a background consisting of BRB80T (BRB80 supplemented with 1 mM GTP or GMPCPP and 100 μM Paclitaxel), for microtubules, and BRB80Ca (BRB80 supplemented with 1 mM GTP or GMPCPP and 1 mM $CaCl_2$) for free GTP-tubulin solutions. Steady-State Fluorescence spectroscopy was performed using a Photon Technology International (PTI) QuantaMaster 40-F NA spectrofluorometer. All steady-state measurements were performed at room temperature.

**Time-Correlated Single Photon Counting (TCPSC)**. The time resolved fluorescence of each sample was measured using a DeltaFlex TCSPC instrument Horiba (Kyoto, Kyoto, Japan). The input light was provided using a 305 nm LED light source (DD300; Horiba, Kyoto, Kyoto, Japan). The usage of 305 nm light ensured that tyrosine and phenylalanine were not photoexcited. A long pass dielectric filter with a transition wavelength of 325 nm was placed after the solution and before the detector to reduce scatter signal. The fluorescence emission was detected at 335 nm. These excitation and detection wavelengths were selected to ensure that tyrosine and phenylalanine were not photoexcited. The time resolved spectra were corrected for instrument response function (IRF) using an aqueous Ludox colloidal silica solution (Fig. S3a). The measurement range was set from 0-100 ns, with the recording set to stop measurements when a maximum of 10,000 counts was attained. Obtained time resolved fluorescence decays were fit to equation (S1):

$$I(t) = A_1 e^{-T_1 t} + A_2 e^{-T_2 t} + A_3 e^{-T_3 t} \quad (S1)$$

The fit parameters $T_1$ and $T_2$ represented tryptophan lifetimes in the protein, while $T_3$ represented the fit parameter for residual scattering from different tubulin polymers, and $A_1$, $A_2$ and $A_3$ represented the relative contributions of each lifetime (amplitude). Average weighted values were estimated by finding the weighted lifetime of tryptophan fluorescence lifetime (the parameters $T_1$, $T_2$ and $T_3$ were weighted using amplitudes $A_1$, $A_2$ and $A_3$) and then determining the average of this value over $n = 3$ to 5 experiments.

**Negative Stain Electron Microscopy.** 2 μM tubulin was polymerized as described as above, then either used at 2 μM or diluted immediately prior to grid preparation to 400 nM. Dilution was performed in the appropriate polymerization buffer at room temperature. 3 μL tubulin mixture was added to glow-discharged CF400-Cu grids (Electron Microscopy Sciences, Hatfield, PA, USA) and stained with 0.75% uranyl formate. Samples were imaged using a Thermo Scientific Talos L120C transmission electron microscope operating at 200 keV. The nominal magnification was 72,000, and defocus values ranged from -0.5 to -2.0 μm. Micrographs were recorded on a Thermo Scientific Ceta-M 4k x 4k-pixel CMOS camera using the TIA data collection software (Thermo Scientific, Waltham, MA, USA), at a calibrated pixel size of 2.02 Å. All TCSPC experiments were performed at room temperature.

**Acknowledgements.** This work was supported by the Templeton World Charity Forum (Project ID: TWCF0530). T.J.A.C acknowledges funding from the U.S. Army Research Office, Department of Defense, under contract no. W911NF-19-1-0373-(74884-PH) and from the Nova Southeastern University President's Faculty Research Development grant (PFRDG334807). The authors thank Dr. Kyu Hyung Park, Dr. Junwoo Kim (Chungbuk National University, Korea) and Prof. Aristide Dogariu (University of Central Florida, FL, USA) for useful comments during the preparation of this manuscript. The authors thank Dr. Gary Laevsky at the Princeton Confocal Microscopy Facility for assisting with fluorescence microscopy. The authors acknowledge the use of Princeton's Imaging and Analysis Center (IAC), which is partially supported by the Princeton Center for Complex Materials (PCCM), a National Science Foundation (NSF) Materials Research Science and Engineering Center (MRSEC; DMR-2011750). A.P.K dedicates this work to his father, Prof. Prem Kumar Kalra.

### Author Contributions.

A.P.K, A.B, D.G.O, T.J.A.C, S.R.H, M.B.M, J.A.T, S.P, R.P and G.D.S designed research, A.P.K, A.B, S.M.T, E.A.Z and A.M.S performed the research, M.B.M, J.A.T, S.P and G.D.S provided new reagents/analytic tools, A.P.K and A.B analyzed the data, A.P.K, A.B and G.D.S wrote the paper.

### Competing Interest Statement

The authors declare no competing interest

## Supplementary Information Appendix for

Electronic Energy Migration in Microtubules

Aarat P. Kalra [a], Alfy Benny [a], Sophie M. Travis [b], Eric A. Zizzi [c], Austin Morales-Sanchez [a], Daniel G. Oblinsky [a], Travis J. A. Craddock [d], Stuart R. Hameroff [e], M. Bruce MacIver [f], Jack A. Tuszyński [c, g, h], Sabine Petry [b], Roger Penrose [i], Gregory D. Scholes [a, *]

[a] Department of Chemistry, New Frick Chemistry Building, Princeton University, NJ 08544, USA
[b] Department of Molecular Biology, Schultz Laboratory, Princeton University, NJ 08544, USA
[c] Department of Mechanical and Aerospace Engineering (DIMEAS), Torino 10129, Italy
[d] Departments of Psychology & Neuroscience, Computer Science, and Clinical Immunology, Nova Southeastern University, Ft. Lauderdale, FL 33314, USA
[e] Department of Anesthesiology, Center for Consciousness Studies, University of Arizona, Tucson, Arizona, USA
[f] Department of Anesthesiology, Stanford University School of Medicine, Stanford, CA 94305, USA
[g] Department of Physics, University of Alberta, Edmonton, Alberta T6G 2E1, Canada
[h] Department of Oncology, University of Alberta, Edmonton, Alberta T6G 1Z2, Canada
[i] Mathematical Institute, Andrew Wiles Building, University of Oxford, Radcliffe Observatory Quarter, Woodstock Road, Oxford, OX2 6GG, United Kingdom

[*] To whom correspondence should be addressed: gscholes@princeton.edu

# DOCKING OF ETOMIDATE, ISOFLURANE AND PICROTOXIN TO HUMAN TUBULIN

**Homology Modelling.** We built homology models of the human tubulin dimer starting from the corresponding amino acid sequences of the alpha-1A and beta-1 chains obtained from UniProt (accession codes Q71U36 and P07437 respectively) (1). The central dimer of the pdb structure 3J6F corresponding to a sheet of a GDP-bound dynamic microtubule with a resolution of 4.90 Å was used as a template (2). Homology models were build using the Homology Modelling tool of the MOE software (3), and the best model was chosen based on the GB/VI score (4). Highly fluctuating C-terminal tails were excluded from the obtained model for all subsequent analyzes. The quality of the generated model was assessed by comparing the phi-psi plot and the QMeanDisCo score between the model and the 3J6F template (5).

**Structure Equilibration.** The obtained homology model was completed by adding the $Mg^{2+}$ ion and the GTP and GDP ligands in their experimental binding positions as found in the template, after adjusting their protonation state using the Protonate3D tool in MOE at 300K and pH=7.4. The tubulin dimer was parametrized using the AMBER99SB-ILDN force field (6), while ligands were parametrized using GAFF (7) and their partial charges calculated using the AM1-BCC method (8). Energy minimization, equilibration and production MD runs were carried out using GROMACS 2021.4 (9). More in detail, the final model of the human tubulin dimer was energy minimized for 1000 steps, and subsequently equilibrated for 100 ps in the NVT ensemble using the v-rescale thermostat with the C-alpha atoms position-restrained with a force of 1000 kcal/mol/nm (10). We then carried out a further 500-ps equilibration in the NPT ensemble using the v-rescale thermostat and the Berendsen barostat with the same position restraints as before (11). We finally lifted the restraints and performed a production molecular dynamics run of 200 ns for further extended structural equilibration in the NTP ensemble using the v-rescale thermostat and Parrinello-Rahman barostat (12). We regarded the first 100 ns of this production MD simulation as additional extended structure equilibration, and we extracted the dominant conformation states of the dimer as the centroids after clustering the last 100 ns of this MD production run. The cluster tool provided by GROMACS was used, with the cutoff value set to 0.15 nm. The subsequent docking simulations were run using these centroids as the target structure.

**Site identification and docking.** After determining the dominant conformations of the dimer from the MD ensemble, we performed a scan for potential binding sites on the dimer using the SiteFinder tool provided in MOE. To focus the analysis not only on the most likely binding sites, but also on pockets with a potentially lower affinity, we investigated all predicted binding sites with a positive PLB score (13), and docked the four different ligands within these regions. In addition, to further complement the analysis, the ligands were also docked to three experimentally known and clinically relevant binding sites on the tubulin dimer, namely the (a) Colchicine binding site, located near the inter-monomer interface and defined by loop βT7, helix βH8, strands βS8 and βS9 and the αT5 loop; (b) the taxane binding site, located on the β subunit surrounded by helix H1, the H6-H7 loop, the H7 helix, the M loop and the S9-S10 loop; this site is partially overlapping with site number 5 found by the SiteFinder tool (c) the vinca alkaloid binding site, at the longitudinal dimer-dimer interface within protofilaments, surrounded by loop αT7, helix αH10 and strand αS9.

Docking runs were carried out in MOE, with the following methodology: we performed the initial placement of the ligands using the triangle matcher algorithm, which describes the active site using α spheres. The generated poses were then scored using the London dG scoring function (3). Subsequently, for each ligand, the 30 best poses found using the triangle matcher algorithm were further energy-minimized starting from their initial placement, using a Molecular Mechanics approach with the protein kept rigid. The final poses were finally re-scored using the GBVI/WSA dG scoring function (14), and the top 5 poses were considered for analysis.

**Results**

**(A) Quality of the Homology Models.** The human alpha 1A and beta 1 tubulin chains shared a sequence identity of 99.3% and 97.9% with the corresponding chains of the 3J6F template, respectively, thus ensuring the reliability of the homology modelling approach. Indeed, on the phi-psi plot, the generated human alpha/beta dimer model featured 94.18% of residues in core regions, 5.35% in allowed regions and 0.47% outliers. In comparison, the same figures for the 3J6F template were 96.81%, 2.83% and 0.36% respectively. The QMeanDisCo global score of the model was 0.76 ± 0.05, which is comparable within error to the corresponding score of the crystallographic template (0.79 ± 0.05) (see Fig. S12: comparison of QMean scores). Both quality assessments confirm the comparable structural quality of the homology model with respect to its 3J6F template.

**(B) Docking.** Clustering of the last 100 ns of the MD ensemble at a 0.15 nm RMSD cutoff yielded a single cluster, suggesting overall conformational stability and no major rearrangements in the protein structure. The centroid of this cluster, representing the dominant conformational state of the tubulin dimer throughout the simulation, was chosen as the conformation to perform docking against.

First, we scanned for possible binding sites on the tubulin dimer using the SiteFinder tool in MOE, and found a total of 43 putative binding sites, with PLB scores ranging from 3.66 (best) to -0.79 (worst). All binding sites with positive PLB values were kept as docking sites, resulting in a total of 13 analyzed binding sites. A summary of these binding sites is provided in Table S6 together with the experimentally known binding sites on tubulin for colchicine, taxanes and vinca alkaloids. In the following, the binding site numbering of Table S6 will be used to identify the sites.

**Etomidate** Predicted binding energies for Etomidate range from -6.77 to -4.36 kcal/mol (mean±std:-5.71±0.51 kcal/mol). The best binding energy was obtained on site 5 (-6.77 kcal/mol), and the interaction between Etomidate and the protein in this site is highlighted in Fig. S9a below. The average results in all 16 binding sites are summarized in Table S7.

**Isoflurane** Predicted binding energies for Isoflurane range from -4.64 to -3.66 kcal/mol (mean±std:-4.11±0.22 kcal/mol), indicative of a slightly weaker predicted interaction with respect to etomidate. The best binding energy was obtained on site 3 (-4.64 kcal/mol), and the corresponding interaction between isoflurane and the protein in this site is highlighted in Fig. S9b, although it is to be underlined how the differences between the different binding poses and binding sites are realistically all within the error of the docking methodology. The average results in all 16 binding sites are summarized in Table S8.

**Picrotoxinin** Predicted binding energies for Picrotoxinin range from -5.88 to 0.56 kcal/mol (mean±std: 4.22±1.79 kcal/mol). The best binding energy was obtained on the taxol binding site (-5.88/kcal/mol), and the corresponding interaction between the ligand and the protein in this site is highlighted in Fig. S9c. The average results in all 16 binding sites are summarized in Table S9.

**Picrotin** Predicted binding energies for Picrotin range from -5.72 to -2.82 kcal/mol (mean±std:-4.89±0.64 kacl/mol). The best binding energy was obtained in the taxol binding site (-5.72 kcal/mol), and the corresponding interaction between the ligand and the protein in this site is highlighted in Fig. S9d. The average results in all 16 binding sites are summarized in Table S10.

**Summary of docking studies.** Overall, for all compounds except isoflurane, the highest predicted affinity is consistently found to be in or near (site 5) the taxol binding site, which is a known and relevant binding site on the tubulin dimer. Predicted energies in the other putative binding locations are nevertheless similar and most are within the thermal noise level of 0.6 kcal/mol. Hence, no hard conclusions can be drawn regarding a distinctive preferred binding site for either of the analyzed ligands on human tubulin.

Picrotoxin is a 1:1 molar mixture of picrotin and picrotoxinin, two highly similar compounds. Hence, docking was performed for both ligands, and no substantial difference in predicted binding affinities emerged, coherently with their high chemical similarity. No conclusions can be drawn from the docking simulations to distinguish a more active compound on tubulin among the two; in fact, both compounds showed the best predicted dG value in the taxol binding site, where they are predicted to form at least two hydrogen bonds with residues ARG318, ARG359 and ARG276 with differences well within the error of docking predictions.

### KINETIC MONTE CARLO SIMULATIONS

**Preparation of microtubule crystal structure** We constructed the all-atom structure of the full microtubule using the following protocol. Firstly, the tubulin sheet of PDB structure 7SJ7 (15) was preprocessed using MOE (3) to fix missing residues in the structure, assign the correct protonation states and perform energy minimization to relief atomic clashes. We then extracted the central tubulin dimer of this structure and used it to construct an initial template of a single tubulin ring, composed of 13 dimers, by fitting 13 copies of the dimer onto the original electron density map deposited in the Electron Microscopy Data Bank (accession code EMD-25156), using the "*Fit in Map*" tool of UCSF ChimeraX (16).

To construct the final microtubule, we first carried out 200-ns Molecular Dynamics simulations of the tubulin sheet to equilibrate the structure and obtain a diversified structural ensemble. In detail, simulations were carried out in GROMACS 2021.4 (9) in the NPT ensemble, using the AMBER 99SB-ILDN force field (6), PME electrostatics with a cutoff of 1.2 nm, the velocity rescale thermostat (10) at 300K and the Parrinello-Rahman barostat (12). We subsequently clustered the last 150 ns of this MD simulation with a cutoff of 0.10 nm using the *gmx cluster* tool, and extracted the centroids of each cluster, focusing only on the central dimer of the tubulin sheet. These centroids represent a set of dominant conformations of the tubulin dimer within a microtubule, and were used to build the final microtubule. In greater detail, we generated each of the final tubulin rings by randomly choosing centroids from the MD clusters and RMSD-fitting them onto the template tubulin ring described above. This was repeated for each of the 31 tubulin rings composing the final microtubule, which was assembled by spacing the rings axially by 8.15 nm, as described in earlier literature for the 13:3B MT lattice (17).

In summary, we eventually obtained a final B-lattice microtubule with a length of 252.65 nm and composed of an ensemble of different tubulin dimer conformations, extracted from all-atom MD simulations and representative of the conformational heterogeneity of tubulin within a microtubule.

**Simulation protocol for homo- and hetero-transfer among tryptophan and tyrosine residues.**
The monomers of TYR and TRP were optimized in the CAM-B3LYP/AUG-CC-PVDZ level of theory and the optimized geometry was used to calculate the transition dipole moments using EOM-CCSD/6-311+G(d). The calculations were carried out in Gaussian 16 program package (18). The obtained transition dipole moments were reoriented at the respective molecular coordinates for calculating the long-range coulombic coupling strengths in the 31-long dimer microtubule crystal structure. We took the dissipation rate ($k_{Dis}$) as $3.06 \times 10^8$ s$^{-1}$ (19).

Charge transfer Integral calculations were carried out on 5 representative TYR dimers selected from the microtubule crystal structure with Inter-aromatic ($R_{IA}$) distances below 6.3 Å. A constrained optimization (B3LYP/6-311+G(d)) were performed on the selected TYR dimers by freezing the carbon skeleton prior to charge transfer integral calculations. The charge transfers Integral calculations (B3LYP/6-311+G(d)) were carried out using the dimer projection (DIPRO) method (20) using the CATNIP program package (21) available from https://github.com/JoshuaSBrown/QC_Tools).

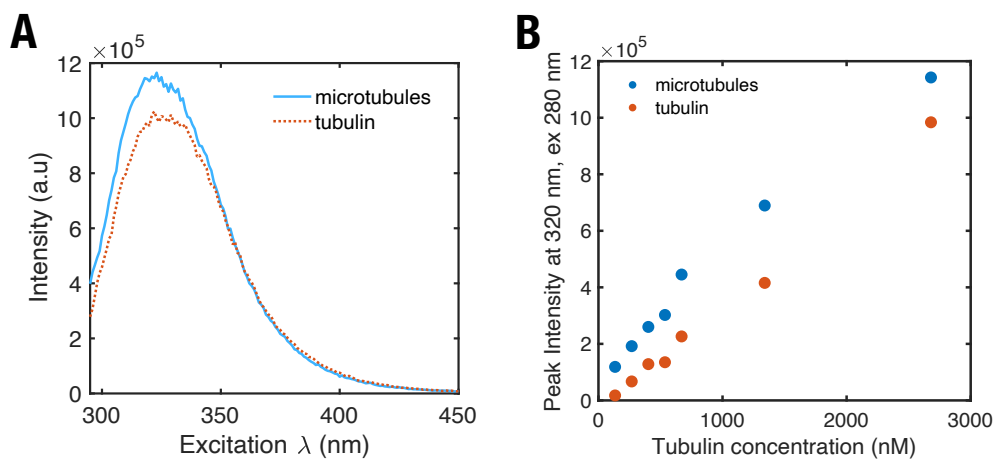

**Fig. S1.** Emission profiles of tubulin and microtubules. (A), The emission spectrum of tubulin and microtubules as a function of concentration, (B), Variation of peak fluorescence intensity at different tubulin concentrations. All samples were excited using λexcitation = 280 nm.

**Electronic Energy Migration in Microtubules** | Preprint

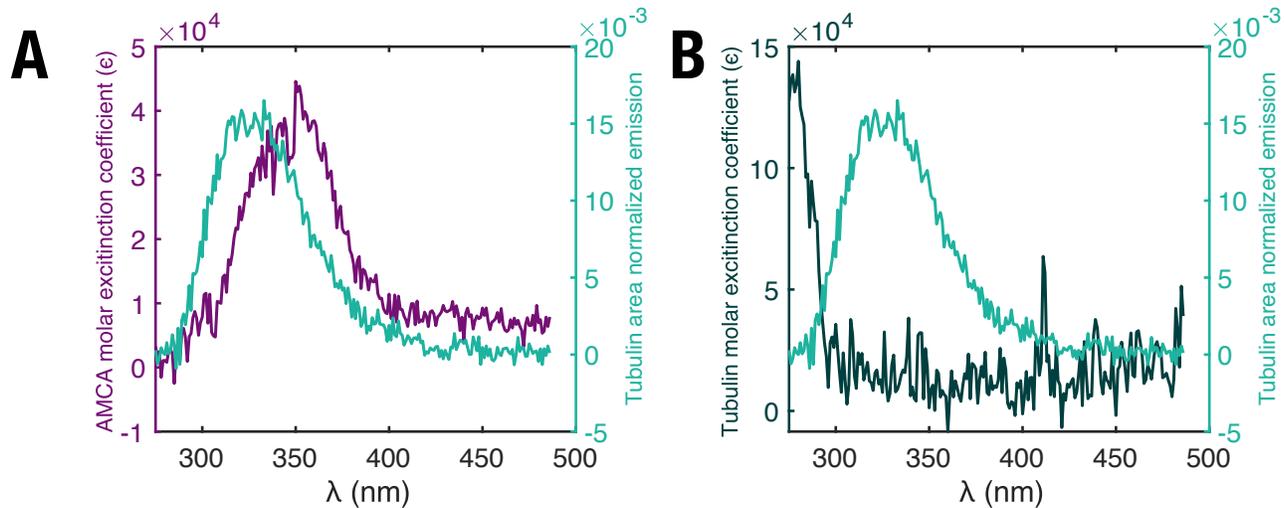

**Fig. S2.** Molar extinction coefficient and area normalized emission spectra of (A) tubulin tryptophan and AMCA, (B) tubulin tryptophan for inter-tryptophan energy transfer.

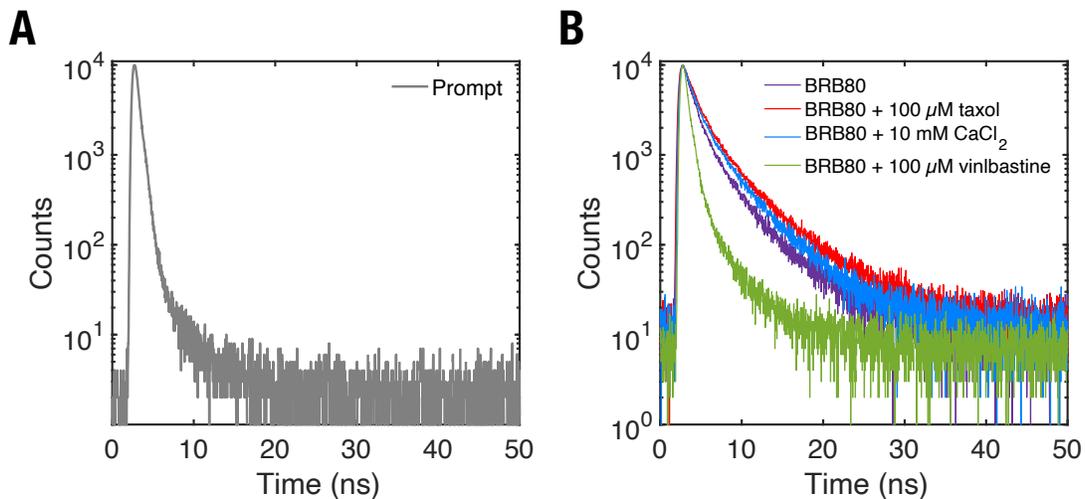

**Fig. S3.** (A), Representative Instrument Response Function (IRF; 'prompt') signal extracted from a ludox colloidal solution, (B), Representative signal from a BRB80, BRB80T (background for microtubule containing solutions), BRB80Ca (background for free GTP-tubulin containing solutions), and BRB80V (background for GTP-tubulin oligomer containing solutions). BRB80V had characteristic lifetimes of 3.67 ns (10.14%), 0.85 ns (5.22%) and 53.6 ps (84.64%).

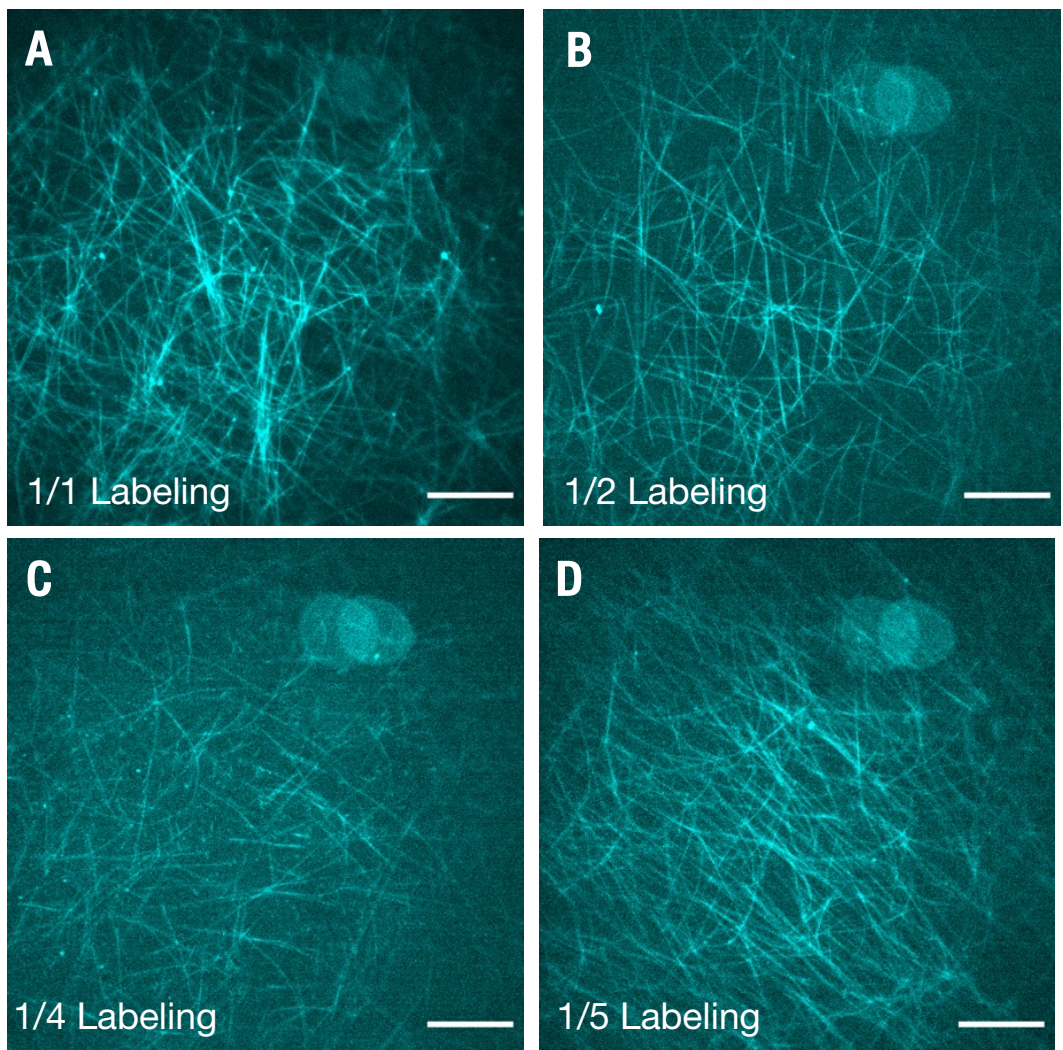

**Fig. S4. Experimentally evaluating the polymerization of microtubules with differing AMCA concentrations using confocal fluorescence microscopy.** Microtubules polymerized using (A), 1/1 labeling ratio (AMCA concentration 2.27 µM in solution; tubulin concentration 2.27 µM in solution), (B), 1/2 labeling ratio (AMCA concentration 1.135 µM), (C), 1/4 labeling ratio (AMCA concentration 0.57 µM) and (D), 1/5 labeling ratio (AMCA concentration 0.45 µM). Scale bars represent 10 µm.

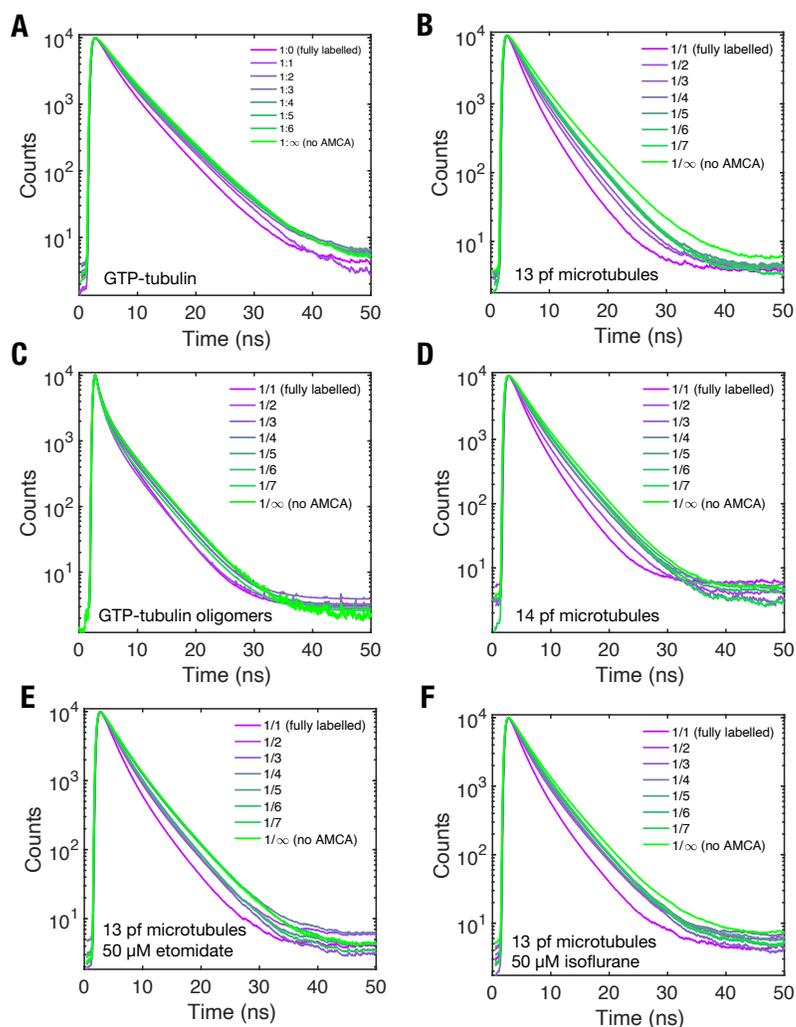

**Fig. S5.** Time resolved fluorescence spectroscopy showing the influence of AMCA labeling ratio on tryptophan fluorescence lifetimes in (A), free GTP-tubulin in solution, (B), 13 protofilament microtubules, (C), GTP-tubulin oligomers, (D), 14 protofilament microtubules (E), 13 protofilament microtubules in the presence of 50 µM etomidate, (F), free 13 protofilament microtubules in the presence of 50 µM isoflurane, as a function of AMCA labeling ratio.

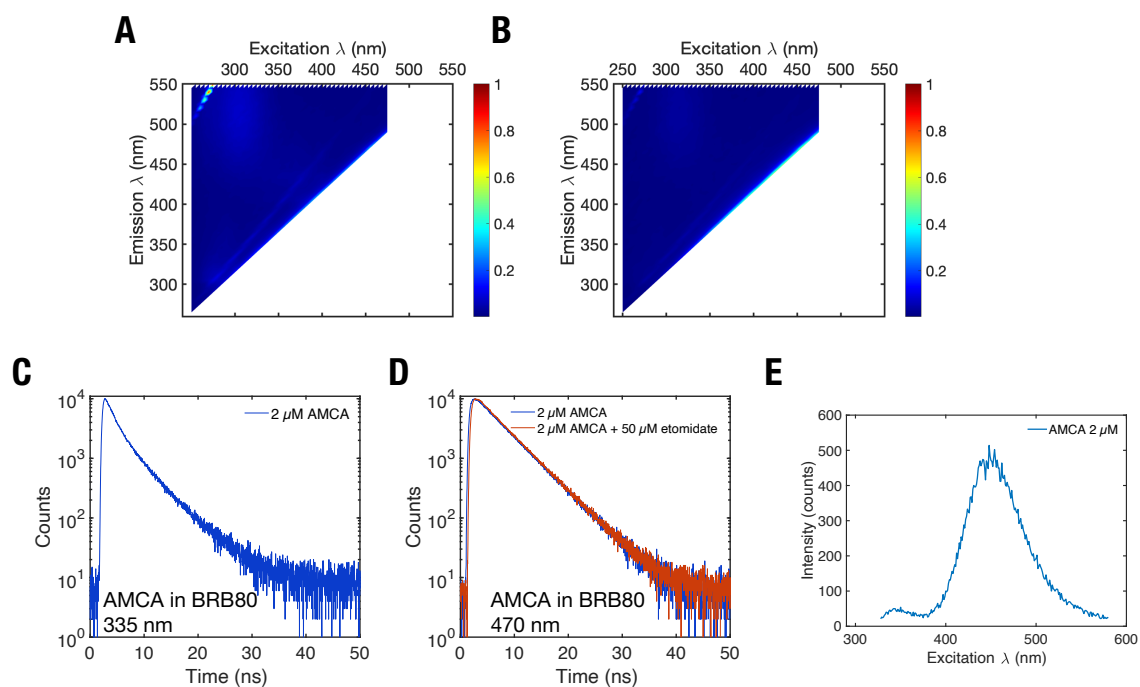

**Fig. S6.** Intensity normalized fluorescence spectra of (A), 50 µM etomidate in BRB80 showing highest fluorescence emission at Raman scattering peaks of water, (B), 50 µM isoflurane in BRB80 showing highest fluorescence emission at Raman scattering peaks of water. These spectra show that etomidate and isoflurane contribute to negligible fluorescence in the excitation 270-310 nm and emission 300-350 nm range, where tryptophan is active. (C), Representative signal from a 2 µM AMCA solution, measured at excitation 305 nm, emission 335 nm (conditions and background solution condition identical to that of GTP-tubulin polymerized microtubules), (D), 470 nm in the presence of agents etomidate, isoflurane and picrotoxin, (E), Steady-state emission spectrum of AMCA at excitation 305 nm recorded using the TCSPC instrumentation.

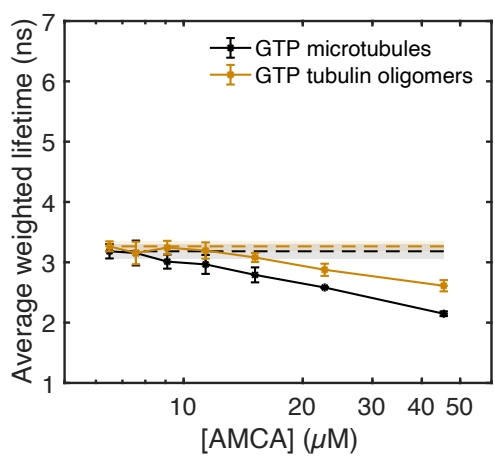

**Fig. S7.** Average weighted tryptophan fluorescence lifetime of GTP tubulin oligomers as a function of AMCA concentration.

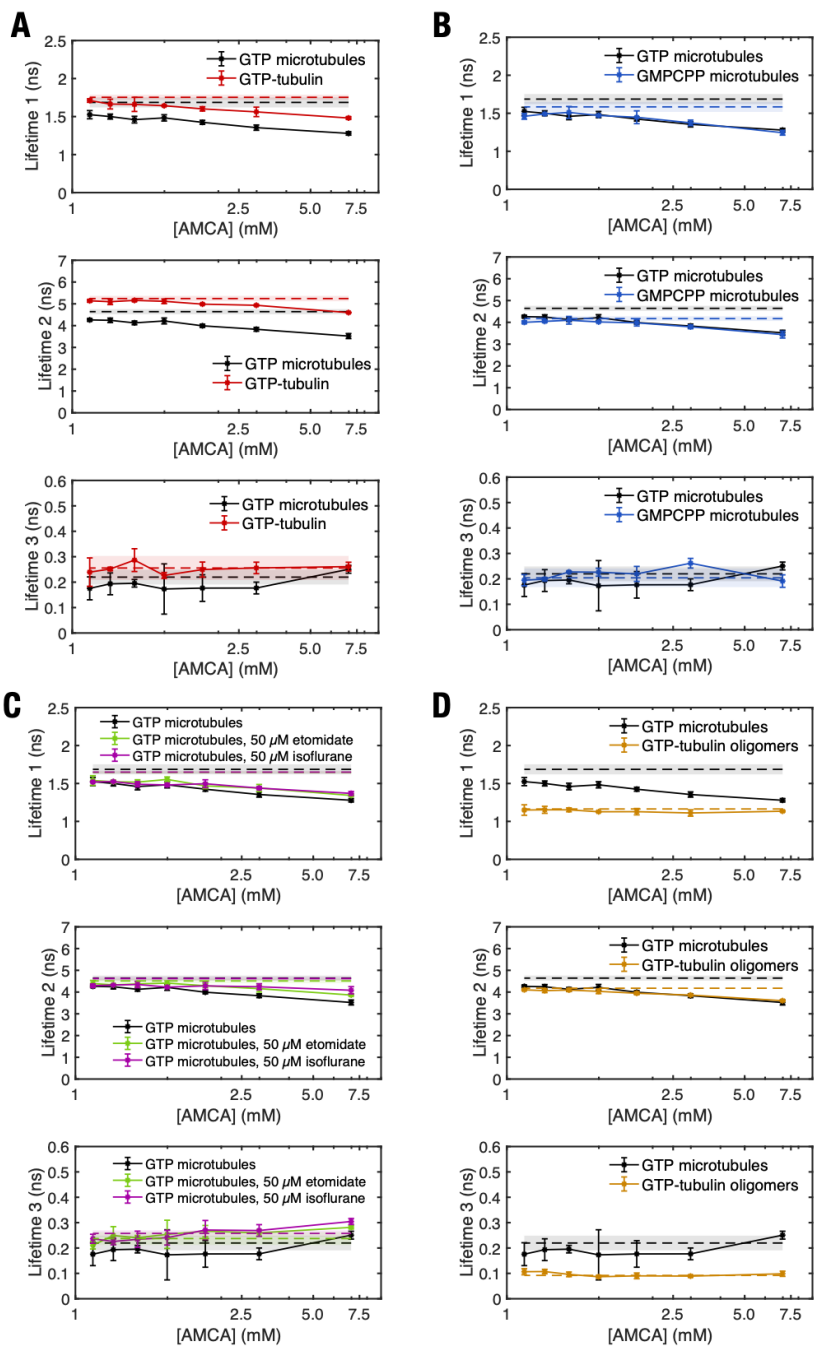

**Fig. S8.** Tryptophan fluorescence lifetimes of different tubulin polymers as a function of AMCA concentration Fitted lifetimes of tryptophan fluorescence as a function of AMCA labeling ratio for (A), GTP-tubulin compared to that of 13 protofilament microtubules, (B), 14 protofilament microtubules compared to that of 13 protofilament microtubules, (C), GTP-tubulin oligomers compared to that of 13 protofilament microtubules, (D), 13 protofilament microtubules in the presence of etomidate and isoflurane compared to that of 13 protofilament microtubules in their absence.

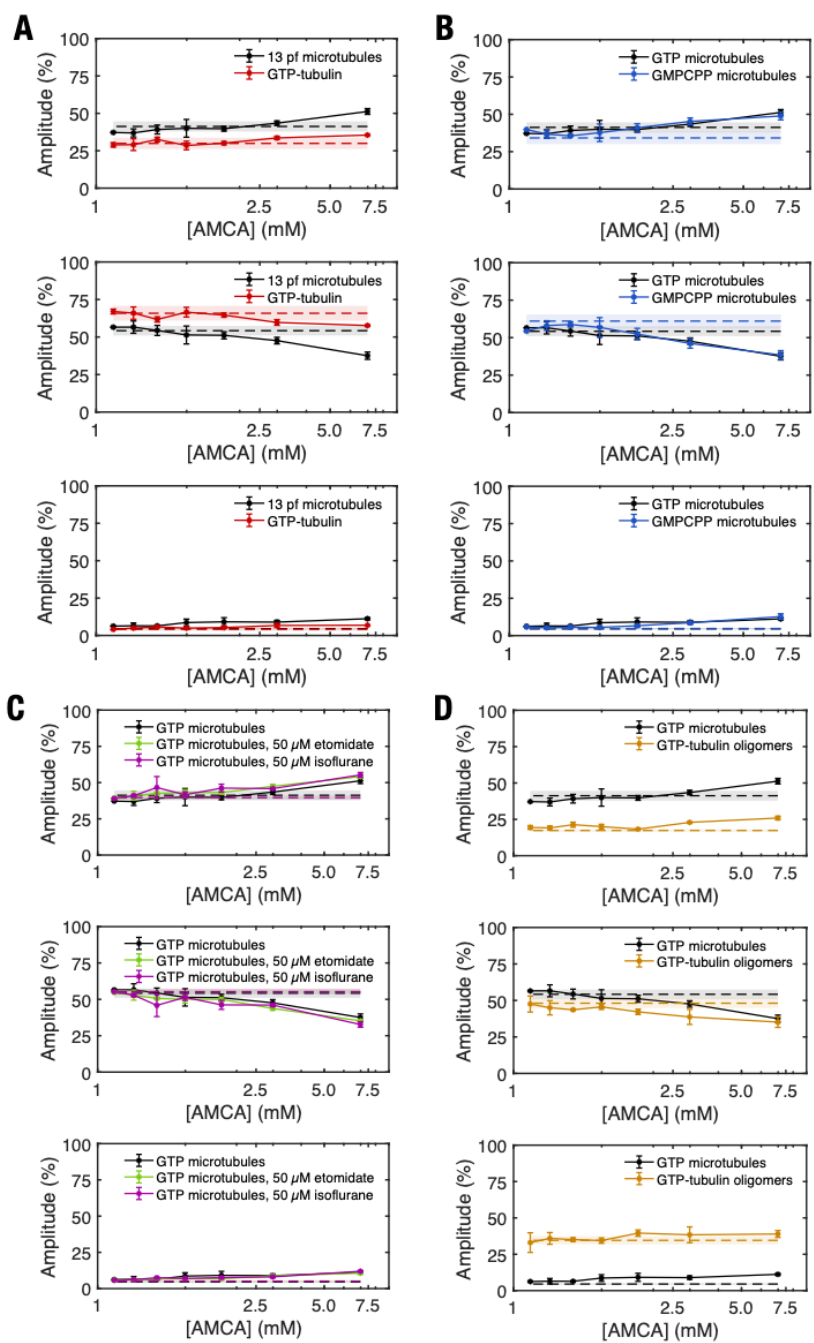

**Fig. S8.** Relative contributions of each lifetime component in tryptophan fluorescence for different tubulin polymers as a function of AMCA concentration. Fitted amplitudes of tryptophan fluorescence as a function of AMCA labeling ratio for (A), GTP-tubulin compared to that of 13 protofilament microtubules, (B), 14 protofilament microtubules compared to that of 13 protofilament microtubules, (C), GTP-tubulin oligomers compared to that of 13 protofilament microtubules, (D), 13 protofilament microtubules in the presence of etomidate and isoflurane compared to that of 13 protofilament microtubules in their absence.

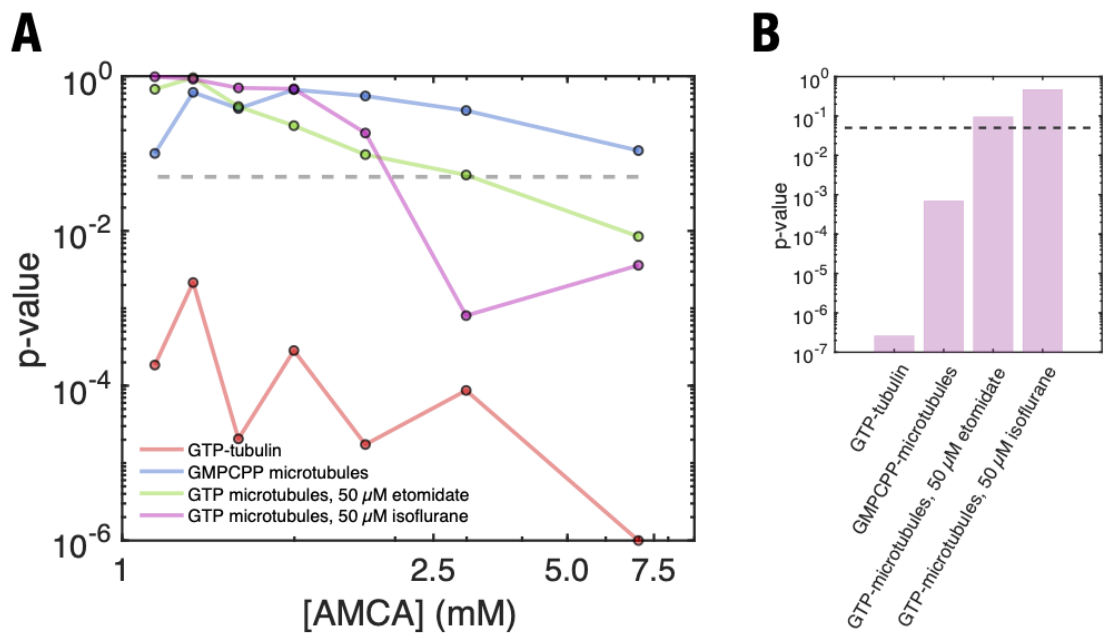

**Fig. S10.** (A) p-values obtained from a two tailed t-test, comparing the values of average weighted lifetimes obtained for GTP-microtubules, to those of GTP-tubulin, GMPCPP microtubules, GTP microtubules in the presence of 50 µM etomidate and GTP microtubules in the presence of 50 µM isoflurane in the presence of varying concentrations of AMCA. The dashed grey line represents a p-value of 0.05. (B) p-values obtained by comparing the average weighted lifetimes obtained for unlabeled microtubules with those of unlabeled GTP-tubulin, unlabeled GMPCPP microtubules, unlabeled GTP microtubules in the presence of 50 µM etomidate and unlabeled GTP microtubules in the presence of 50 µM isoflurane. The dashed line represents a p-value of 0.05.

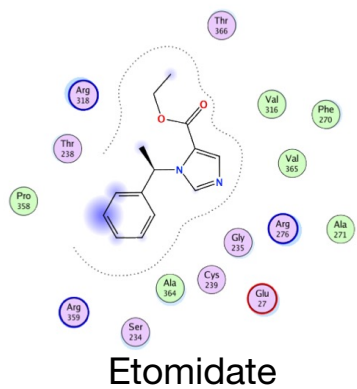
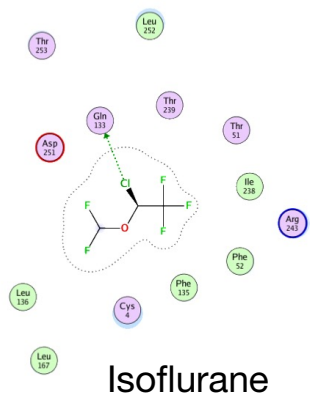
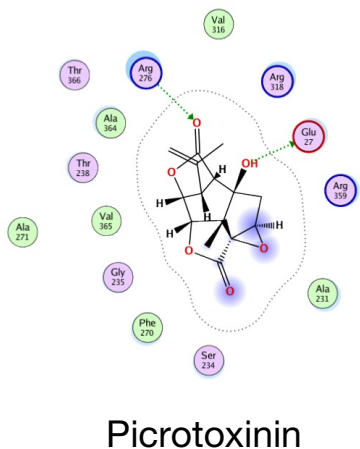
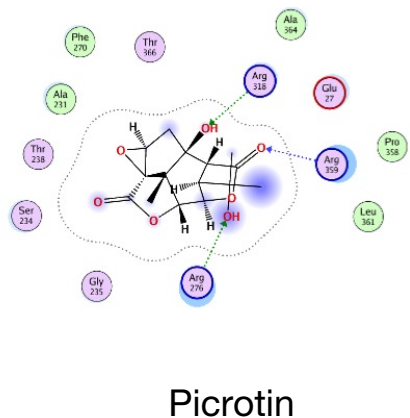

**Fig. S11.** (A), Weak interactions between etomidate and binding site residues of Site 5, which featured the best predicted binding energy, (B), Detailed interactions between isoflurane and binding site residues of Site 3, which featured the best predicted binding energy, (C), Interactions between picrotoxinin and the taxol binding site, which featured the best predicted binding energy, d, Interactions between picrotin and binding site residues of the taxol binding site, which featured the best predicted binding energy.

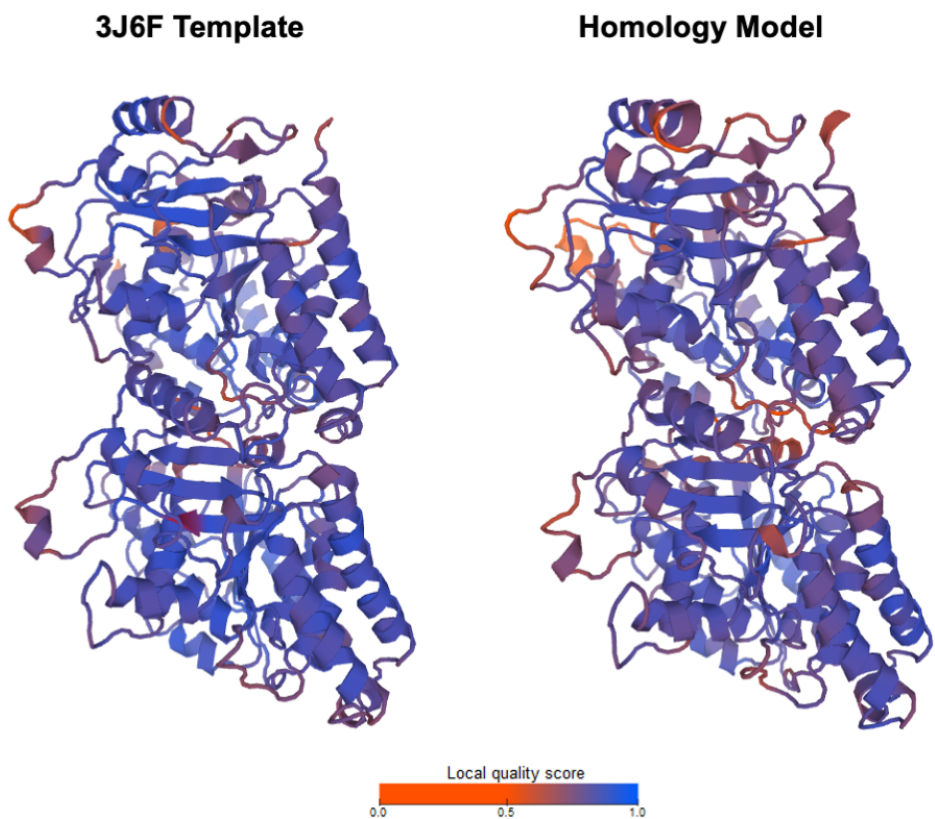

**Fig. S12.** Visual comparison of the distribution of QMeanDisCo scores of the template (PDB ID 3J6F, left) and the final homology model (right).

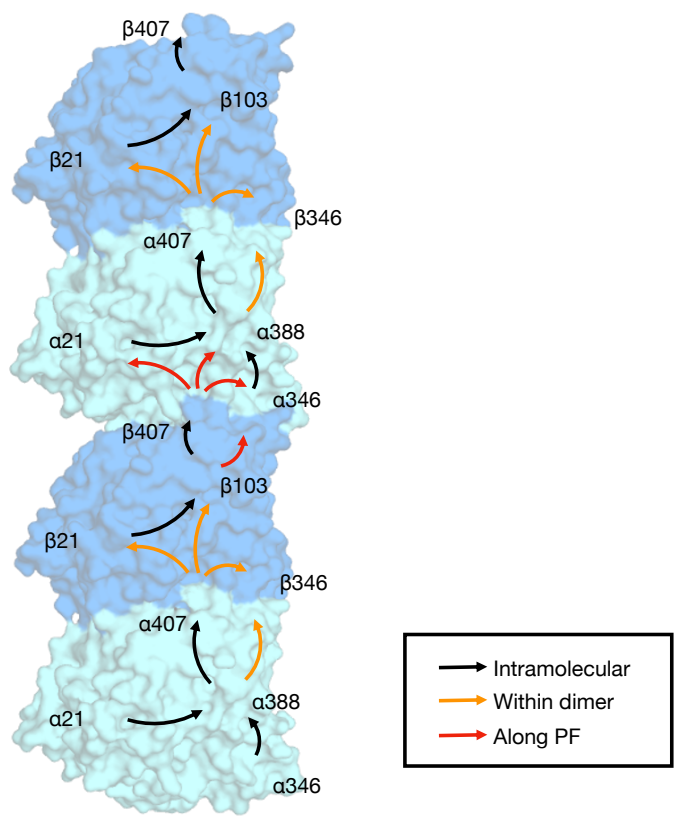

**Fig. S13.** Schematic showing long-range energy transport along a microtubule.

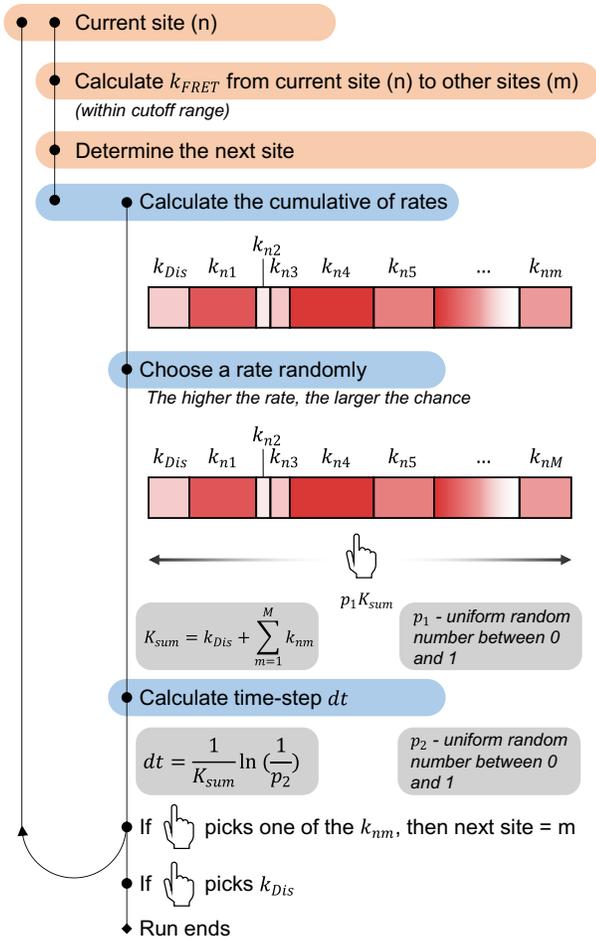

**Fig. S14.** Algorithm used for kinetic Monte Carlo simulations.

## A

Absorption and emission spectra    Spectral overlap

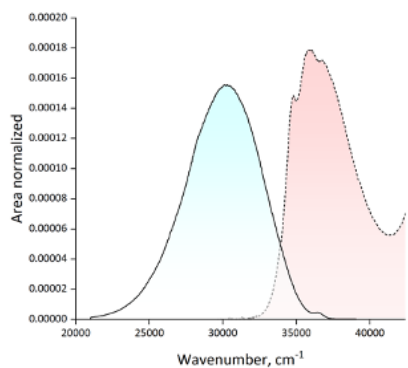
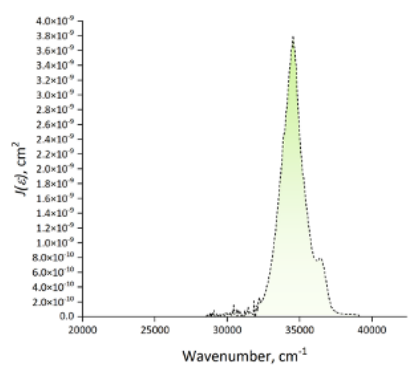

## B

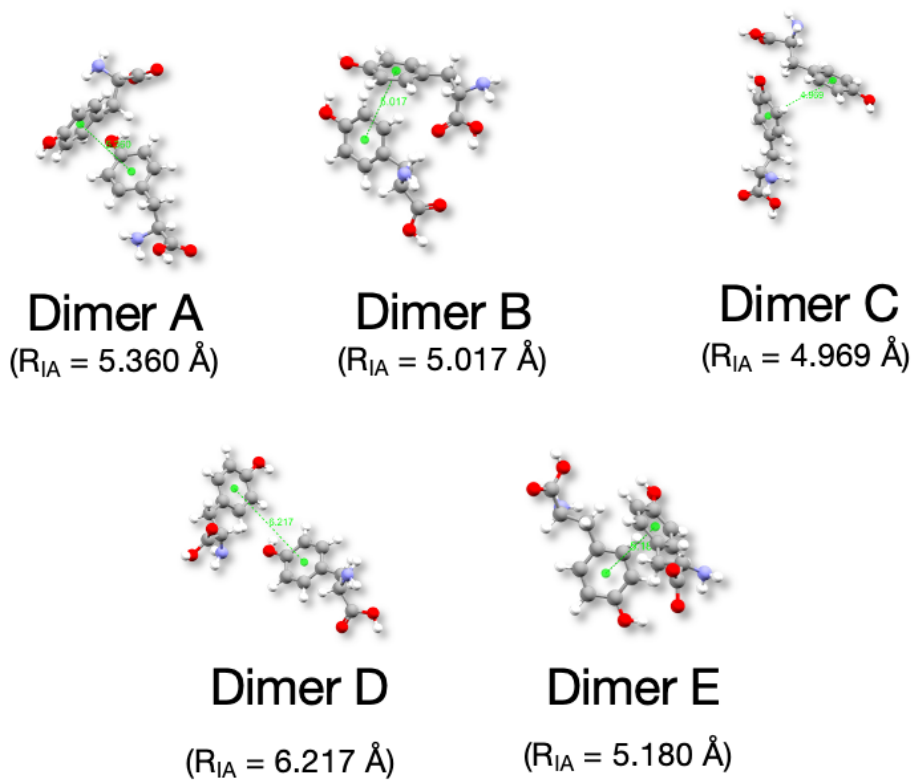

Dimer A  
($R_{IA}$ = 5.360 Å)

Dimer B  
($R_{IA}$ = 5.017 Å)

Dimer C  
($R_{IA}$ = 4.969 Å)

Dimer D  
($R_{IA}$ = 6.217 Å)

Dimer E  
($R_{IA}$ = 5.180 Å)

**Fig. S15.** (A) Spectral overlap integral values for photoexcitation donor acceptor pairs (B) Graphs showing absorption and emission spectra used for calculations. (C) Representative TYR dimers with short inter-aromatic distance ($R_{IA}$) selected for calculating hole and electron transfer integrals.

**Table S1.** Nearest neighbor distances (Å) between tryptophan residues in different tubulin polymers. CD2 (carbon at the delta-2 position) atoms in pairs of tryptophan residues were used to measure distances between residues.

|  | 14 protofilament microtubules (6DPU) | GTP-tubulin (3J6G) | 13 protofilament microtubules (3J6G) | GTP-tubulin oligomers (Vinblastine stabilized) |
|---|---|---|---|---|
| TUBα1 W21 | 27.9 | 28 | 27.7 | 26.6 |
| TUBα1 W346 | 14.3 | 19.9 | 15.2 | 13.8 |
| TUBα1 W388 | 17.9 | 19.9 | 19.9 | 20.3 |
| TUBα1 W407 | 15.8 | 15.9 | 15.9 | 15.9 |
| TUBβ W21 | 26.8 | 26.3 | 26.3 | 26.2 |
| TUBβ W103 | 13.9 | 14.3 | 14.3 | 13.5 |
| TUBβ W346 | 15.8 | 15.9 | 15.9 | 15.9 |
| TUBβ W407 | 13.9 | 14.3 | 14.3 | 13.5 |
| average min | 18.2875 | 19.3125 | 18.6875 | 18.2125 |
| std min | 5.7543617 | 5.325528143 | 5.435974745 | 5.51735897 |

**Table S2.** Distances (Å) between tryptophan residues for 13 and 14 protofilament microtubules.

| Residue 1 | Residue 2 | 14 protofilament microtubule (6DPU) CD2 Distance (Å) | 13 protofilament microtubule (3J6G) CD2 Distance (Å) |
|---|---|---|---|
| **TUBα1 Intradimer** | | | |
| TUBα1 W21 | TUBα1 W346 | 36.8 | 37.6 |
| TUBα1 W21 | TUBα1 W388 | 29.2 | 28 |
| TUBα1 W21 | TUBα1 W407 | 35.4 | 35 |
| TUBα1 W346 | TUBα1 W388 | 17.9 | 19.9 |
| TUBα1 W346 | TUBα1 W407 | 42.8 | 43.5 |
| TUBα1 W388 | TUBα1 W407 | 25.8 | 24.4 |
| **TUBβ Intradimer** | | | |
| Tubβ W21 | Tubβ W103 | 26.8 | 26.3 |
| Tubβ W21 | Tubβ W346 | 38 | 37.9 |
| Tubβ W21 | Tubβ W407 | 35.3 | 35.2 |
| Tubβ W103 | Tubβ W346 | 33.8 | 33.5 |
| Tubβ W103 | Tubβ W407 | 13.9 | 14.3 |
| Tubβ W346 | Tubβ W407 | 45.7 | 45.5 |
| **TUBα1 to TUBβ across dimer** | | | |
| TUBα1 W21 | Tubβ W21 | 41 | 40.8 |
| TUBα1 W21 | Tubβ W103 | 59.3 | 58.3 |
| TUBα1 W21 | Tubβ W346 | 43.3 | 43 |
| TUBα1 W21 | Tubβ W407 | 71.9 | 71.4 |
| TUBα1 W346 | Tubβ W21 | 65.9 | 65.7 |
| TUBα1 W346 | Tubβ W103 | 72.3 | 70.7 |
| TUBα1 W346 | Tubβ W346 | 41.7 | 40.8 |
| TUBα1 W346 | Tubβ W407 | 85.5 | 84.3 |
| TUBα1 W388 | Tubβ W21 | 49.2 | 46.5 |
| TUBα1 W388 | Tubβ W103 | 72.3 | 51.5 |
| TUBα1 W388 | Tubβ W346 | 25 | 23 |
| TUBα1 W388 | Tubβ W407 | 67.8 | 65.1 |
| TUBα1 W407 | Tubβ W21 | 27.7 | 27.7 |
| TUBα1 W407 | Tubβ W103 | 26.9 | 27.7 |
| TUBα1 W407 | Tubβ W346 | 15.8 | 15.9 |
| TUBα1 W407 | Tubβ W407 | 42 | 41.7 |
| **Tubα1 to TUBβ along PROTOFILAMENT** | | | |
| TUBα1 W21 | Tubβ W21 | 42.9 | 42.7 |
| TUBα1 W21 | Tubβ W103 | 37.9 | 38 |
| TUBα1 W21 | Tubβ W346 | 68.1 | 67.5 |
| TUBα1 W21 | Tubβ W407 | 27.9 | 27.7 |
| TUBα1 W346 | Tubβ W21 | 68.1 | 43.9 |
| TUBα1 W346 | Tubβ W103 | 20.7 | 22 |
| TUBα1 W346 | Tubβ W346 | 42.3 | 42.7 |
| TUBα1 W346 | Tubβ W407 | 14.3 | 15.2 |
| TUBα1 W388 | Tubβ W21 | 52.8 | 53.9 |
| TUBα1 W388 | Tubβ W103 | 33.4 | 35.7 |
| TUBα1 W388 | Tubβ W346 | 60 | 61.8 |
| TUBα1 W388 | Tubβ W407 | 20.4 | 22.1 |
| TUBα1 W407 | Tubβ W21 | 72.9 | 72.4 |
| TUBα1 W407 | Tubβ W103 | 55.8 | 55.8 |
| TUBα1 W407 | Tubβ W346 | 85.6 | 85.3 |
| TUBα1 W407 | Tubβ W407 | 42 | 41.6 |
| **Tubα1 to TUBα1 across lattice** | | | |
| TUBα1 W21 | TUBα1 W21 | 48.6 | 48.6 |
| TUBα1 W21 | TUBα1 W346 | 79.2 | 79.8 |
| TUBα1 W21 | TUBα1 W388 | 71.9 | 70.3 |
| TUBα1 W21 | TUBα1 W407 | 63.7 | 63.6 |
| TUBα1 W346 | TUBα1 W21 | 48.4 | 50.4 |
| TUBα1 W346 | TUBα1 W346 | 61.4 | 62.3 |
| TUBα1 W346 | TUBα1 W388 | 58.6 | 59.4 |
| TUBα1 W346 | TUBα1 W407 | 57.5 | 58.4 |
| TUBα1 W388 | TUBα1 W21 | 48.1 | 48.7 |
| TUBα1 W388 | TUBα1 W346 | 67.4 | 68.3 |
| TUBα1 W388 | TUBα1 W388 | 59.9 | 59.5 |
| TUBα1 W388 | TUBα1 W407 | 50.6 | 50.7 |
| TUBα1 W407 | TUBα1 W21 | 64.5 | 64.1 |
| TUBα1 W407 | TUBα1 W346 | 88.6 | 87.6 |
| TUBα1 W407 | TUBα1 W388 | 77.1 | 74.8 |
| TUBα1 W407 | TUBα1 W407 | 58.9 | 58.8 |
| **Tubβ to TUBβ across lattice** | | | |
| Tubβ W21 | Tubβ W21 | 48.9 | 48.6 |
| Tubβ W21 | Tubβ W103 | 60.4 | 59.7 |
| Tubβ W21 | Tubβ W346 | 81.2 | 81.1 |
| Tubβ W21 | Tubβ W407 | 63.7 | 63.1 |
| Tubβ W103 | Tubβ W21 | 59.6 | 59.4 |
| Tubβ W103 | Tubβ W103 | 59 | 58.7 |
| Tubβ W103 | Tubβ W346 | 84.9 | 85 |
| Tubβ W103 | Tubβ W407 | 60.9 | 59.7 |
| Tubβ W346 | Tubβ W21 | 47.8 | 47.2 |
| Tubβ W346 | Tubβ W103 | 48.3 | 47 |
| Tubβ W346 | Tubβ W346 | 62 | 61.8 |
| Tubβ W346 | Tubβ W407 | 56.7 | 55.6 |
| Tubβ W407 | Tubβ W21 | 64.4 | 64.5 |
| Tubβ W407 | Tubβ W103 | 60.9 | 60.9 |
| Tubβ W407 | Tubβ W346 | 90.4 | 90.6 |
| Tubβ W407 | Tubβ W407 | 58.6 | 58.4 |

**Table S4.** Estimation of 2D diffusion length using Stern-Volmer analysis.

| Variable | Symbol | Value | Estimation method |
|---|---|---|---|
| Stern-Volmer quenching constant | $k_Q$ | See Table 1 | TCSPC analysis |
| Tryptophan weighted average lifetime in tubulin polymer in the absence of AMCA | $\tau_0$ | See Fig. 4 | TCSPC |
| Tryptophan weighted average lifetime in tubulin polymer as a function of [AMCA] | $\tau$ | See Fig. 4 | TCSPC |
| sum of the radii of tryptophan and AMCA | $r$ | 1 nm | approximate sum |

**Table S5.** Comparison of 2D diffusion lengths as estimated by Stern-Volmer kinetics and RET using Förster theory.

| Tubulin polymerization state | $L_{SV}$ (nm) |
|---|---|
| Free GTP-tubulin | 4.07 ± 0.1 |
| 13 protofilament microtubules (GTP-tubulin polymerized) | 6.64 ± 0.1 |
| 14 protofilament microtubules (GMPCPP-tubulin polymerized) | 6.13 ± 0.1 |
| GTP-tubulin oligomers | 4.63 ± 0.1 |
| 13 protofilament microtubules with 50 µM etomidate | 5.61 ± 0.1 |
| 13 protofilament microtubules with 50 µM isoflurane | 5.81 ± 0.2 |

**Table S6.** Summary of all binding sites found by SiteFinder with a PLB score greater than 0, together with the known binding sites on tubulin of colchicine, taxol and vinca alkaloids.

| Site Number | Subunit | Residues lining the site | PLB score |
|---|---|---|---|
| 1 | A | ARG2 GLU27 HIS28 LYS40 GLY45 ASP46 SER48 PHE49 ASN50 THR51 PHE52 ALA240 LEU242 ARG243 PHE244 ASP245 ASN249 ASN356 GLN358 | 3.66 |
| 2 | A/B | GLN134 ASP197 GLU198 TYR200 VAL236 THR237 LEU240 LEU250 LEU253 ALA254 MET257 PHE266 ILE368 | 2.83 |
| 3 | A | THR191 HIS192 THR193 THR194 LEU195 GLU196 ARG264 ILE265 PHE267 ASP424 LEU428 | 2.15 |
| 4 | A/B | GLN15 THR73 VAL74 ASP76 GLU77 THR80 TYR224 LEU42 ASP45 ARG46 PRO243 GLY244 GLN245 ARG320 MET321 SER322 ASP355 | 1.68 |
| 5 | B | ALA19 GLU22 LEU23 LEU26 ARG229 VAL362 VAL363 GLY366 ASP367 LEU368 | 1.45 |
| 6 | B | HIS107 THR150 SER151 LEU152 MET154 GLU155 THR193 THR194 GLU196 HIS197 SER198 | 1.18 |
| 7 | A | PHE255 ASN258 LEU259 PRO261 MET313 ALA314 CYS315 CYS316 TRP346 CYS347 PRO348 GLY350 PHE351 LYS352 | 1.10 |
| 8 | B | ARG2 GLU27 HIS28 LYS40 GLY45 ASP46 SER48 PHE49 ASN50 THR51 PHE52 ALA240 LEU242 ARG243 PHE244 ASP245 ASN249 ASN356 GLN358 | 0.38 |
| 9 | A | GLN134 ASP197 GLU198 TYR200 VAL236 THR237 LEU240 LEU250 LEU253 ALA254 MET257 PHE266 ILE368 | 0.35 |
| 10 | A/B | THR191 HIS192 THR193 THR194 LEU195 GLU196 ARG264 ILE265 PHE267 ASP424 LEU428 | 0.22 |
| 11 | A | GLN15 THR73 VAL74 ASP76 GLU77 THR80 TYR224 LEU42 ASP45 ARG46 PRO243 GLY244 GLN245 ARG320 MET321 SER322 ASP355 | 0.20 |
| 12 | A | ALA19 GLU22 LEU23 LEU26 ARG229 VAL362 VAL363 GLY366 ASP367 LEU368 | 0.06 |
| 13 | A | HIS107 THR150 SER151 LEU152 MET154 GLU155 THR193 THR194 GLU196 HIS197 SER198 | 0.04 |
| Colchicine | A/B | αGLU71 αASN101 βARG241 βARG251 βARG318 βALA352 | - |
| Taxol | B | VAL23 ASP26 LEU217 THR219 VAL229 SER230 MET233 PRO272 THR274 SER275 ARG276 GLY277 SER278 | - |
| Vinca | A | ALA247 LEU248 ASN249 LYS352 VAL353 | - |

**Table S7.** Predicted GBVI/WSA binding energy of etomidate in each analyzed binding site, reported as mean ± standard deviation among the top 5 poses in each site after energy minimization and rescoring.

| Site | GBVI/WSA ΔG [kcal/mol] |
|---|---|
| 1 | -6.10 ± 0.18 |
| 2 | -5.45 ± 0.25 |
| 3 | -5.22 ± 0.12 |
| 4 | -5.76 ± 0.13 |
| 5 | -6.53 ± 0.23 |
| 6 | -5.51 ± 0.15 |
| 7 | -5.18 ± 0.04 |
| 8 | -5.85 ± 0.14 |
| 9 | -5.34 ± 0.08 |
| 10 | -6.22 ± 0.29 |
| 11 | -4.85 ± 0.22 |
| 12 | -5.29 ± 0.05 |
| 13 | -5.48 ± 0.21 |
| Colchicine | -5.56 ± 0.23 |
| Taxol | -6.15 ± 0.09 |
| Vinca | -5.90 ± 0.21 |

**Table S8.** Predicted GBVI/WSA binding energy of isoflurane in each analyzed binding site, reported as mean ± standard deviation among the top 5 poses in each site after energy minimization and rescoring.

| Site | GBVI/WSA ΔG [kcal/mol] |
|---|---|
| 1 | -4.21 ± 0.06 |
| 2 | -4.15 ± 0.16 |
| 3 | -4.52 ± 0.09 |
| 4 | -4.27 ± 0.07 |
| 5 | -4.21 ± 0.15 |
| 6 | -3.83 ± 0.09 |
| 7 | -3.93 ± 0.10 |
| 8 | -3.81 ± 0.18 |
| 9 | -3.94 ± 0.07 |
| 10 | -4.19 ± 0.10 |
| 11 | -3.94 ± 0.07 |
| 12 | -3.91 ± 0.05 |
| 13 | -4.21 ± 0.11 |
| Colchicine | -4.22 ± 0.03 |
| Taxol | -4.07 ± 0.10 |
| Vinca | -4.14 ± 0.15 |

**Table S9.** Predicted GBVI/WSA binding energy of Picrotoxinin in each analyzed binding site, reported as mean ± standard deviation among the top 5 poses in each site after energy minimization and rescoring.

| Site | GBVI/WSA ΔG [kcal/mol] |
| --- | --- |
| 1 | -5.03 ± 0.09 |
| 2 | -5.26 ± 0.09 |
| 3 | -4.89 ± 0.03 |
| 4 | -5.11 ± 0.29 |
| 5 | -5.54 ± 0.11 |
| 6 | -5.03 ± 0.12 |
| 7 | -4.92 ± 0.05 |
| 8 | -4.82 ± 0.18 |
| 9 | -4.93 ± 0.07 |
| 10 | -5.00 ± 0.33 |
| 11 | -4.60 ± 0.14 |
| 12 | -4.87 ± 0.15 |
| 13 | -0.40 ± 1.12 |
| Colchicine | -1.16 ± 0.71 |
| Taxol | -5.57 ± 0.20 |
| Vinca | -0.46 ± 1.04 |

Table S10. Predicted GBVI/WSA binding energy of picrotin in each analyzed binding site, reported as mean ± standard deviation among the top 5 poses in each site after energy minimization and rescoring.

| Site | GBVI/WSA ΔG [kcal/mol] |
|---|---|
| 1 | -5.23 ± 0.11 |
| 2 | -5.3 ± 0.24 |
| 3 | -5.08 ± 0.09 |
| 4 | -5.36 ± 0.17 |
| 5 | -5.55 ± 0.11 |
| 6 | -4.94 ± 0.15 |
| 7 | -4.94 ± 0.11 |
| 8 | -5.03 ± 0.1 |
| 9 | -4.86 ± 0.29 |
| 10 | -5.32 ± 0.1 |
| 11 | -4.46 ± 0.17 |
| 12 | -4.8 ± 0.3 |
| 13 | -3.84 ± 0.42 |
| Colchicine | -4.29 ± 1.12 |
| Taxol | -5.58 ± 0.12 |
| Vinca | -3.68 ± 0.58 |

**Table S11.** Spectral overlap integrals between tyrosine and tryptophan residues.

| D-A pair | $J_{SO}$ ($10^{-6}$ cm) |
|---|---|
| Tryptophan > Tryptophan | 7.53 |
| Tyrosine > Tyrosine | 16.15 |
| Tryptophan > Tyrosine | 5.63 |
| Tyrosine > Tryptophan | 30.52 |

**Table S12.** Magnitudes of hole ($|t_h|$) and electron transfer integrals ($|t_e|$) for TYR dimers at short inter-aromatic distances.

|  | $R_{IA}$(Å) | $|t_h|$ (cm$^{-1}$) | $|t_e|$ (cm$^{-1}$) |
|---|---|---|---|
| **Dimer A** | 5.36 | 49.62 | 34.83 |
| **Dimer B** | 5.02 | 12.62 | 85.56 |
| **Dimer C** | 4.97 | 52.25 | 137.47 |
| **Dimer D** | 6.22 | 51.25 | 175.40 |
| **Dimer E** | 5.18 | 89.53 | 419.39 |

**SI References**